\documentclass[%
 aip,
 amsmath,amssymb,
preprint,%
]{revtex4-1}
\usepackage{graphicx}
\usepackage[export]{adjustbox}
\usepackage{caption}
\usepackage{dcolumn}
\usepackage{bm}
\usepackage[version=4]{mhchem}
\usepackage{color}
\usepackage[utf8]{inputenc}
\usepackage[decimalsymbol=comma,expproduct=\cdot]{siunitx}
\usepackage[utf8]{inputenc}
\usepackage{amsmath}
\usepackage{amssymb}
\usepackage{hyperref}
\usepackage[retainorgcmds]{IEEEtrantools}
\usepackage[mathscr]{euscript}
\usepackage{mathtools}
\usepackage{mhchem}
\usepackage{siunitx}
\usepackage{enumitem}
\usepackage{array,booktabs}
\newcolumntype{C}{>{$\displaystyle}c<{$}} 
\usepackage{subfigure}

\newcommand{\eqnref}[1]{(\ref{#1})}

\begin{document}
\preprint{AIP/123-QED}
\title[Jo \textit{et al.}]{Rovibrational-Specific QCT and Master Equation Study on $\text{N}_2(\text{X}^1\Sigma_g^+)$+O$({}^3\text{P})$ and NO$(\text{X}^2\Pi)$+N$({}^4\text{S})$ Systems in High-Energy Collisions}
\author{Sung Min Jo}
\affiliation{Center for Hypersonics and Entry Systems Studies (CHESS), \\ University of Illinois at Urbana-Champaign, Urbana, IL 61801, USA}
\author{Simone Venturi}
\affiliation{Center for Hypersonics and Entry Systems Studies (CHESS), \\ University of Illinois at Urbana-Champaign, Urbana, IL 61801, USA}
\author{Maitreyee P. Sharma}
\affiliation{Center for Hypersonics and Entry Systems Studies (CHESS), \\ University of Illinois at Urbana-Champaign, Urbana, IL 61801, USA}
\author{Alessandro Munaf\`{o}}
\affiliation{Center for Hypersonics and Entry Systems Studies (CHESS), \\ University of Illinois at Urbana-Champaign, Urbana, IL 61801, USA}
\author{Marco Panesi}
\email{mpanesi@illinois.edu}
\affiliation{Center for Hypersonics and Entry Systems Studies (CHESS), \\ University of Illinois at Urbana-Champaign, Urbana, IL 61801, USA}
\date{\today}

\begin{abstract}
This work presents a detailed investigation of the energy transfer and dissociation mechanisms in $\text{N}_2(\text{X}^1\Sigma_g^+)$+O$({}^3\text{P})$ and NO$(\text{X}^2\Pi)$+N$({}^4\text{S})$  systems using rovibrational-specific quasi-classical trajectory (QCT) and master equation analyses. The complete set of state-to-state kinetic data, obtained via QCT, allows for an in-depth investigation of the Zel'dovich mechanism leading to the formation of NO molecules at microscopic and macroscopic scales. The master equation analysis demonstrates that the low-lying vibrational states of $\text{N}_2$ and NO have dominant contributions to the NO formation and the corresponding extinction of $\text{N}_2$ through the exchange process. For the considered temperature range, it is found that while nearly 50\% of the dissociation processes for $\text{N}_2$ and NO occurs in the molecular quasi-steady-state (QSS) regime, the amount of the Zel'dovich reaction is zero. Using the QSS approximation to model the Zel'dovich mechanism leads to an overestimation of NO production by more than a factor of 4 in the high-temperature range. The breakdown of this well-known approximation has profound consequences for the approaches that heavily rely on the validity of QSS assumption in hypersonic applications. The investigation of the rovibrational state population dynamics reveals substantial similarity among different chemical systems for the energy transfer and the dissociation processes, providing promising physical foundations for the use of reduced-order strategies to other chemical systems without significant loss of accuracy. 
\end{abstract}

\maketitle

\section{Introduction}\label{sec:intro}
In hypersonic flows, the formation and extinction of nitric oxide (NO) are important since this chemical species emits strong ultraviolet radiation inside shock layers. One of the most well-known reaction pathways contributing to NO formation in high-temperature air is the following heterogeneous exchange process (\emph{i.e.}, Zel'dovich reaction):
\begin{equation}
\text{N}_2+\text{O} \rightarrow \text{NO}+\text{N}.
\label{eq:NO_Zel'dovich}
\end{equation}
\noindent
Due to the fast dissociation of diatomic oxygen ($\mathrm{O}_2$), collisions between molecular nitrogen ($\mathrm{N}_2$) and atomic oxygen (O) control NO formation through reaction \eqnref{eq:NO_Zel'dovich}. The existing studies \cite{Walch_JCP_1987,Bose_JCP_1996} found that the above reaction predominantly occurs through the lowest triplet surfaces $^3A^{\prime}$ and $^3A^{\prime\prime}$ in the hypersonic flow regime, provided that the non-adiabatic transitions and spin-orbit coupling are neglected. This implies that the colliding species (\emph{e.g.}, $\mathrm{N}_2$ and O) remain in their electronic ground states after the collision.

In addition to the Zel'dovich mechanism \eqnref{eq:NO_Zel'dovich}, collisional energy transfer and dissociation processes of individual collision pairs in the $\text{N}_2$O system (\emph{i.e.}, $\text{N}_2$+O and NO+N) are also relevant in developing a reliable thermochemical non-equilibrium model for high-temperature air. Regarding the $\text{N}_2$O system, previous theoretical studies \cite{Bose_JCP_1996,Gamallo_2003,Luo_2017,Koner_2020,DenisAlpizar_NON,Lin_2016,Lin_2016_QCT} have been carried out by means of \emph{ab-initio} potential energy surface (PES) constructions followed by quasi-classical trajectory (QCT) calculations for particular chemical reaction channels. Each of the PESs developed by Bose and Candler \cite{Bose_JCP_1996}, and Gamallo \emph{et al.} \cite{Gamallo_2003} was employed to compute thermal rate coefficients for the Zel'dovich reaction \eqnref{eq:NO_Zel'dovich}. Luo \emph{et al.} \cite{Luo_2017} computed the same quantity along with the thermal dissociation rate coefficient in $\text{N}_2$+O collisions by relying on the PESs from Gamallo \emph{et al.} \cite{Gamallo_2003}. Koner \emph{et al.} \cite{Koner_2020} improved the accuracy of the PESs of Denis-Alpizar \emph{et al.} \cite{DenisAlpizar_NON} by using neural networks. Lin \emph{et al.} \cite{Lin_2016,Lin_2016_QCT} constructed global PESs that cover a wide range of reactive configurations, and they numerically investigated the reactive trajectories for the Zel'dovich mechanism \eqnref{eq:NO_Zel'dovich}. The aforementioned studies \cite{Bose_JCP_1996,Gamallo_2003,Luo_2017,Koner_2020,DenisAlpizar_NON,Lin_2016,Lin_2016_QCT} focused on specific portions of the reaction dynamics that can occur on the $\text{N}_2$O surfaces, instead of the overall energy transfer and dissociation mechanisms. In addition, to the authors' best knowledge, no detailed state-to-state (StS) master equation analysis has been carried out for the $\text{N}_2$O system. This inhibits a deeper understanding of the non-equilibrium chemical kinetics in high-temperature air.

By means of rovibrational StS master equation analyses, it is possible to investigate detailed chemical-kinetic processes in terms of rovibrational state population dynamics, which provide a crucial component to the development of reduced-order models.
In particular, the study of rovibrational distributions from the $\text{N}_2(\text{X}^1\Sigma_g^+)$+N$({}^4\text{S})$ \cite{PANESI_2013_BOXRVC,Heritier_JCP_2014,Panesi_2014_Shock} and $\text{O}_2(\text{X}^3\Sigma_g^-)$+O$({}^3\text{P})$ \cite{Venturi2020} systems have inspired various coarse-graining strategies \cite{Munafo_EPJD_2012,SAHAI_ADAPTIVE,Sharma_2020,Venturi2020} for model reduction in non-equilibrium chemistry.
Although those approaches were developed based on strong physical foundations and refused \emph{ad hoc} assumption, they have been only tested for $\text{N}_2(\text{X}^1\Sigma_g^+)$+N$({}^4\text{S})$ \cite{Munafo_EPJD_2012,SAHAI_ADAPTIVE,Sharma_2020}, $\text{N}_2(\text{X}^1\Sigma_g^+)$+$\text{N}_2(\text{X}^1\Sigma_g^+)$ \cite{MACDONALD_MEQCT,MACDONALD_vs_DMS}, $\text{O}_2(\text{X}^3\Sigma_g^-)$+O$({}^3\text{P})$ \cite{Venturi2020}, $\text{CO}({}^1\Sigma^+$)+$\text{O}({}^3\text{P})$ \cite{VENTURI_CO2_CGM}, and $\text{CO}_2(\text{X}^1\Sigma_g^+)$+M \cite{Sahai_2019_PRF} (M denotes inert species) chemical systems. This aspect motivates a systematic comparative investigation on the similarities among different chemical systems and the general applicability of the coarse-graining strategies, with the ultimate goal of constructing a comprehensive reduced-order approach for high-temperature air.

Toward this end, the present study proposes a detailed investigation on the rovibrational energy transfer and dissociation processes in the complete $\text{N}_2$O molecular system by means of QCT and master equation analyses. A complete set of rovibrational-specific rate coefficients, including the inelastic, dissociation, and homogeneous and heterogeneous exchange processes, are calculated in a wide range of kinetic temperatures. Then, they are employed to integrate a set of master equations for an ideal chemical reactor problem. Macroscopic quantities, such as the quasi-steady-state (QSS) reaction rate coefficient and the internal energy relaxation time, are evaluated using the rovibrational population distributions to compare with existing data. The similarity of internal energy transfer and dissociation among different chemical systems is investigated by comparing the rovibrational population dynamics. The present master equation analysis also aims to understand the mechanism of the Zel’dovich reaction \eqnref{eq:NO_Zel'dovich} on the microscopic scale and to assess the validity of the QSS approximation in modeling that chemical reaction.

This paper is organized as follows: The related PESs, the constructed StS kinetic database, the system of master equations, and the assessment strategy of order reduction technique for the Zel'dovich mechanism are discussed in Sec. \ref{sec:phys}. In Sec. \ref{sec:results}, the results of the investigation are presented and it is divided into four subsections: Sec. \ref{sec:energy-iso} investigates the internal energy transfer processes, and Sec. \ref{sec:all-iso} focuses on the analysis of the dissociation-recombination dynamics. Section \ref{sec:all} provides the detailed investigation on mechanism of the Zel'dovich reaction. In Sec. \ref{sec:rate}, the present macroscopic rate coefficients are compared with existing data from literature. Finally, Sec. \ref{sec:conc} provides the conclusions from the present work.

\section{Physical modeling}\label{sec:phys}
\subsection{Potential energy surfaces for $\text{N}_2$O}\label{sec:PES}
The several PESs have been constructed for the $\text{N}_2(\text{X}^1\Sigma_g^+)$+O$({}^3\text{P})$ and NO$(\text{X}^2\Pi)$+N$({}^4\text{S})$ systems\cite{Bose_JCP_1996,Gamallo_2003,Lin_2016,DenisAlpizar_NON,Koner_2020}. The collisions predominantly occur on the lowest triplet surfaces in the $A^{\prime}$ and $A^{\prime\prime}$ symmetry and, therefore, the focus has been placed on developing the $^3A^{\prime}$ and $^3A^{\prime\prime}$ PESs. One of the earliest kinetics study for the Zel'dovich reaction was carried out by Bose and Candler~\cite{Bose_JCP_1996} using the \emph{ab-initio} data from Walch~\emph{et al.}~\cite{Walch_JCP_1987} and Gilibert~\emph{et al.}~\cite{Gilibert_1992_JCP}. This study provided thermal rate coefficients for the exchange reaction \eqnref{eq:NO_Zel'dovich} in the temperature range 3000-8000 K. 
However, the \emph{ab-initio} points were computed using complete active space self-consistent field/contracted configuration interaction (CASSCF/CCI) method. Furthermore, the points were computed only in selected portions of the surface. The low level of theory used to compute the \emph{ab-initio} points and lack of points throughout the entire domain prompted further development of the $^3A^{\prime}$ and $^3A^{\prime\prime}$ surfaces. In 2003, Gamallo~\emph{et al.}~\cite{Gamallo_2003} developed the PESs using second-order perturbation theory on a complete active-space self-consistent-field wave function, CASPT2 method. As a downside, CASPT2 method makes the PESs less reliable for studying dissociation reactions. 
Furthermore, it was found in later studies,\cite{Lin_2016,DenisAlpizar_NON,Koner_2020} that the $^3A^{\prime}$ surface has additional transition states and local minima that were not accounted for in the PES by Gamallo~\emph{et al.} \cite{Gamallo_2003}. Accurate consideration of the stationary points is important since they directly affect the reaction pathways of the system being studied. The PESs developed by Lin~\emph{et al.}~\cite{Lin_2016}, Denis-Alpizar~\emph{et al.}~\cite{DenisAlpizar_NON}, and Koner~\emph{et al.}~\cite{Koner_2020} use \emph{ab-initio} points computed at the multi-reference configuration interaction (MRCI) and MRCI with Davidson's correction (MRCI+Q) level of theory, thereby making them more suitable to study the high-energy collisions of the $\text{N}_2\text{O}$ system. 
Lin~\emph{et al.}~\cite{Lin_2016} used MRCI energies, which were improved using the dynamically scaled external correlation method, and the multi-body component of the PES was then fitted using permutationally invariant polynomials in mixed exponential-Gaussian bond order variables. 
Denis-Alpizar~\emph{et al.}~\cite{DenisAlpizar_NON} and Koner~\emph{et al.}~\cite{Koner_2020} employed the MRCI+Q energies fitted using a reproducing kernel Hilbert space (RKHS) scheme.

For the purpose of the present study, the $^3A^{\prime}$ and $^3A^{\prime\prime}$ PESs developed by Lin~\emph{et al.}~\cite{Lin_2016} are used.
These surfaces were constructed to be adiabatic and neglect spin-orbit coupling. The principal reason for using these PESs is that the range of geometries used in constructing the analytical surface is adequate to accurately study both energy transfer and dissociation. 
They are particularly favorable for representing the highly endothermic forward exchange reaction \eqnref{eq:NO_Zel'dovich}, for which the two reaction pathways on the $^3A^{\prime}$ surface and the one on the $^3A^{\prime\prime}$ surface are characterized by accurately described stationary points. The reverse reaction, NO+N $\rightarrow$ $\mathrm{N}_2$+O, is barrierless on the $^3A^{\prime\prime}$ surface and has a barrier ($\approx 10.5$ kcal/mol) on the $^3A^{\prime}$ surface, making it an exothermic reaction.  
However, the PESs  by Lin \emph{et al.} \cite{Lin_2016} are limited in their accuracy in studying such an exothermic reverse exchange reaction, NO+N $\rightarrow$ $\mathrm{N}_2$+O, since they do not have a complete representation for the weak long-range interactions in the NO+N channel. 
Albeit, as it will be shown later in Sec.~\ref{sec:rate}, in the temperature range of interest to this work, these PESs \cite{Lin_2016} are able to provide a good representation of the energy transfer as well as dissociation-recombination reactions. 

\subsection{State-to-state kinetic database}\label{sec:kinetics}
In this work, the following collisional energy transfer and dissociation-recombination processes are considered by assuming that all components remain in their electronic ground state:
\begin{itemize}[leftmargin=*]
\item Rovibrational excitation and de-excitation through inelastic and homogeneous exchange:
\begin{equation}
\text{NO}\left(i\right)+\text{N} \mathrel{\mathop{\rightleftarrows}^{k_{i \rightarrow j}^{\text{NO}} }_{k_{j \rightarrow i}^{\text{NO}}}} \text{NO}\left(j\right)+\text{N},
\label{eq:NO_Inel-HomoExch}
\end{equation}
\begin{equation}
\text{N}_2\left(m\right)+\text{O} \mathrel{\mathop{\rightleftarrows}^{k_{m \rightarrow l}^{\text{N}_2} }_{k_{l \rightarrow m}^{\text{N}_2}}} \text{N}_2\left(l\right)+\text{O}.
\label{eq:N2_Inel}
\end{equation}
\item Rovibrational dissociation and recombination:
\begin{equation}
\text{NO}\left(i\right)+\text{N} \mathrel{\mathop{\rightleftarrows}^{k_{i \rightarrow c}^{\text{NO}} }_{k_{c \rightarrow i}^{\text{NO}}}} \text{N}+\text{O}+\text{N},
\label{eq:NO_Diss}
\end{equation}
\begin{equation}
\text{N}_2\left(m\right)+\text{O} \mathrel{\mathop{\rightleftarrows}^{k_{m \rightarrow c}^{\text{N}_2} }_{k_{c \rightarrow m}^{\text{N}_2}}} \text{N}+\text{N}+\text{O}.
\label{eq:N2_Diss}
\end{equation}
\item Rovibrational energy transfer through Zel'dovich mechanism:
\begin{equation}
\text{NO}\left(i\right)+\text{N} \mathrel{\mathop{\rightleftarrows}^{k_{i \rightarrow m}^{E,\text{NO}} }_{k_{m \rightarrow i}^{E,\text{N}_2}}} \text{N}_2\left(m\right)+\text{O}.
\label{eq:N2O_HetExch}
\end{equation}
\end{itemize}
\noindent
In Eqs. \eqnref{eq:NO_Inel-HomoExch}-\eqnref{eq:N2O_HetExch}, the $k$ symbol stands for the rate coefficient for the considered process. The index pairs $(i,\, j)$ and $(m, \, l)$ denote the rovibrational states of NO and $\mathrm{N}_2$, respectively. These internal levels are stored by increasing energy in the sets $\mathcal{I}_{\mathrm{NO}}$ and $\mathcal{I}_{\mathrm{N}_2}$. To each rovibrational state there corresponds a unique vibrational ($v$) and rotational ($J$) quantum number (\emph{e.g.}, $i = i(v, \, J)$). The subscript $c$ (\emph{i.e.}, continuum) refers to the dissociated state. Rovibrational states data (\emph{e.g.}, energy levels) are obtained by solving Scr$\ddot{\text{o}}$dinger's equation based on the WKB semi-classical approximation. \cite{Truhlar1979,SCHWENKE_VVTC_1988} As a result of these calculations, $\mathrm{N}_2$ has 9093 levels with $v_{max}$=53, whereas NO has 6739 levels with $v_{max}$=45, including both bound and quasi-bound states.

\begin{figure}[h]
    \centering
    \includegraphics[width=0.75\textwidth]{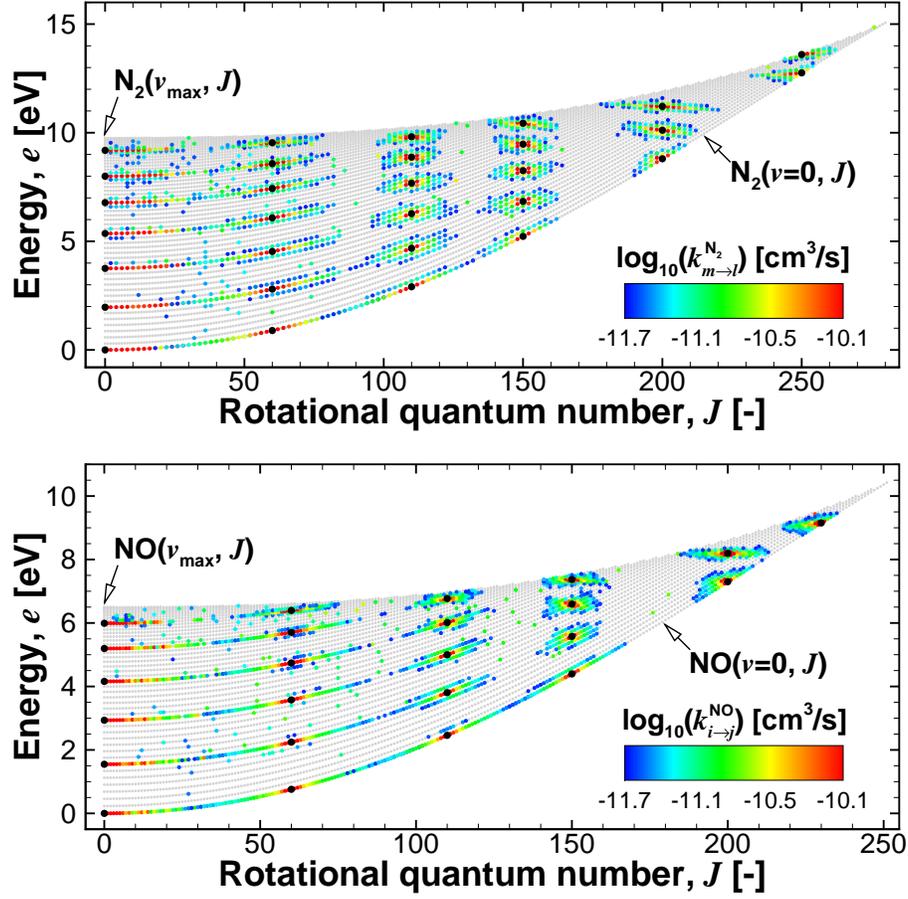}
    \caption{Distributions of the energy transfer rate coefficients in $\text{N}_2$+O (top) and NO+N (bottom) at $T$=\SI{10000}{\kelvin}, overlayed on the diatomic potentials. The gray dots indicate that the rate coefficients corresponding to those final rovibrational levels are below a cutoff value. The black dots represent the initial states.}
    \label{fig:InelExch_Dist}
\end{figure}

To compute the rovibrational specific StS rate coefficients, QCT calculations are carried out using \textsc{CoarseAIR}, \cite{Venturi2020_ML,Venturi2020} an in-house QCT code that is a modernized version of the original \textsc{VVTC} code developed at NASA Ames Research Center by Schwenke.\cite{SCHWENKE_VVTC_1988} The StS kinetic database constructed here covers the temperature range \num{2500}-\SI{20000}{\kelvin}. Figure \ref{fig:InelExch_Dist} shows the distributions of the energy transfer rate coefficients in $\mathrm{N}_2$+O and NO+N systems at \SI{10000}{\kelvin}. The rate coefficients are overlayed on the diatomic potentials. For $\mathrm{N}_2$+O, the inelastic component is shown, whereas, for NO+N, the summation of inelastic and homogeneous exchange rate coefficients is presented. For both systems, the energy transfer occurs with the highest probability to the rovibrational states right next to the initial one (\emph{i.e.}, the black dots). In the low-lying states, most of the transitions occur within the same vibrational strands, whereas the transition pathways are spread across different vibrational levels for the high-lying states. It is interesting to note that the transition patterns for the energy transfer in Fig. \ref{fig:InelExch_Dist} are significantly similar to each other, although they refer to different chemical systems. 

\subsection{Master equations}\label{sec:master}
The StS kinetic database constructed and discussed in Sec. \ref{sec:kinetics} has been used to perform a rovibrational master equation analysis of  $\text{N}_2(\text{X}^1\Sigma_g^+)$+O$({}^3\text{P})$ and NO$(\text{X}^2\Pi)$+N$({}^4\text{S})$ interactions. The system of equations used for this purpose describes the evolution of the various chemical species as a result of inelastic, exchange and dissociation processes, and reads:
\begin{IEEEeqnarray}{rCl}
\frac{dn_i}{dt}&=&\dot{\omega}_i^I-\sum_m^{\text{N}_2}\dot{\omega}_{i,m}^{E,\text{NO}}-\dot{\omega}_i^D, \label{eq:NO_Pop} \\
\frac{dn_m}{dt}&=&\dot{\omega}_m^I+\sum_i^{\text{NO}}\dot{\omega}_{i,m}^{E,\text{NO}}-\dot{\omega}_m^D, \label{eq:N2_Pop}  \\
\frac{dn_{\text{O}}}{dt}&=&\sum_i^{\text{NO}}\dot{\omega}_i^D+\sum_i^{\text{NO}}\sum_m^{\text{N}_2}\dot{\omega}_{i,m}^{E,\text{NO}}, \label{eq:O_Pop} \\
\frac{dn_{\text{N}}}{dt}&=&\sum_i^{\text{NO}}\dot{\omega}_i^D+2\sum_m^{\text{N}_2}\dot{\omega}_m^D-\sum_i^{\text{NO}}\sum_m^{\text{N}_2}\dot{\omega}_{i,m}^{E,\text{NO}}, \label{eq:N_Pop}
\end{IEEEeqnarray}
where $n_X$ denotes the number density of the $X$ species/level and $t$ identifies time. The mass production rates for excitation, $\dot{\omega}^I$, Zel'dovich reaction, $\dot{\omega}^E$, and dissociation, $\dot{\omega}^D$, are evaluated based on the zeroth-order reaction rate theory \cite{Kustova_book,Giov_book}. Their expressions are provided in App. \ref{app:rates}.

The master equations \eqnref{eq:NO_Pop}-\eqnref{eq:N_Pop} are numerically integrated using \textsc{plato} (PLAsmas in Thermodynamic
nOn-equilibrium) \cite{MUNAFO_HEGEL}, a library for non-equilibrium plasmas developed within The Center for Hypersonics and Entry System Studies (CHESS) at University of Illinois at Urbana-Champaign. In solving the master equations using the kinetic database in Sec. \ref{sec:kinetics}, the exothermic rate coefficients are used as reference for the inelastic and homogeneous exchange processes, while the endothermic ones are reconstructed based on the micro-reversibility. For the Zel'dovich mechanism, however, the endothermic rate coefficient (\emph{i.e.}, for $\text{N}_2$+O $\rightarrow$ NO+N) is employed, while the reverse rate coefficients are reconstructed. This is due to the fact that the reverse rate coefficients have higher level of uncertainty (See Fig. \ref{fig:NO-N_Exch} for more details).

The global rate coefficients can be computed over the time domain from the microscopic state-specific rate coefficients and the rovibrational population distributions. The forward and backward global heterogeneous exchange rate coefficients corresponding to Eq. \eqnref{eq:N2O_HetExch} are defined as
\begin{IEEEeqnarray}{rCl}
\tilde{k}^{E,\text{NO}}&=&\dfrac{1}{n_{\mathrm{NO}}} \sum_m^{\text{N}_2}\sum_i^{\text{NO}}n_i k_{i \rightarrow m}^{E,\text{NO}},\quad m \in \mathcal{I}_{\text{N}_2},\quad i \in \mathcal{I}_{\text{NO}},\label{eq:GlobalHetExchRate_NO_To_N2} \\
\tilde{k}^{E,\text{N}_2}&=&\dfrac{1}{n_{\mathrm{N}_2}} \sum_i^{\text{NO}}\sum_m^{\text{N}_2}n_m k_{m \rightarrow i}^{E,\text{N}_2},\quad m \in \mathcal{I}_{\text{N}_2},\quad i \in \mathcal{I}_{\text{NO}}.
\label{eq:GlobalHetExchRate_N2_To_NO}
\end{IEEEeqnarray}
At equilibrium, Eqs. \eqnref{eq:GlobalHetExchRate_NO_To_N2} and \eqnref{eq:GlobalHetExchRate_N2_To_NO} become the thermal heterogeneous exchange rate coefficients as the internal states follow a Maxwell-Boltzmann distribution. The global dissociation rate coefficients of NO and $\mathrm{N}_2$ are defined following the previous work by Panesi \emph{et al.} \cite{PANESI_2013_BOXRVC}

\subsection{Assessment of reduced-order modeling for Zel'dovich mechanism}\label{sec:reduced}
A previous study by Venturi \emph{et al.} proposed an encoding-decoding strategy for assessing the accuracy of reduced-order models in representing a particular channel of the rovibrational-specific kinetics. The approach was based on comparing the results of fully-StS simulations with the ones obtained from master equation studies in which the state-specific rate coefficients for the channel of interest were first grouped and then reconstructed, while all the coefficients for the remaining processes were kept at the StS level. The strategy was there adopted to analyze the effects of coarse-graining the dissociation pathways, and it is employed here to investigate the accuracy of the existing reduced-order models \cite{Park_book,Munafo_EPJD_2012,SAHAI_ADAPTIVE,MACDONALD_MEQCT,Venturi2020} in describing the Zel'dovich mechanism.

For the grouped internal states, the following collisional heterogeneous exchange process is analyzed:
\begin{equation}
\text{NO}\left(p\right)+\text{N} \leftrightarrow \mathrm{N}_2\left(q\right)+\text{O},\quad p \in \mathcal{I}_{\text{NO}},\quad q \in \mathcal{I}_{\text{N}_2},
\label{eq:HeteroExch_Group}
\end{equation}
where $p$ and $q$ denote the grouped states of NO and $\mathrm{N}_2$, respectively. For the $p$-th and $q$-th groups/bins, the equation governing the time rate of change of the grouped population is obtained by summing the master equations over the states within each group (\emph{i.e.}, zeroth-order moment):
\begin{equation}
\frac{dn_p}{dt}=-k_{p{\rightarrow}q}^{E,\text{NO}}n_pn_{\text{N}}+k_{q{\rightarrow}p}^{E,\text{N}_2}n_qn_{\text{O}}.
\label{eq:HeteroExch_p-th_q-th_Bins}
\end{equation}
The grouped Zel'dovich reaction rates $k_{p \rightarrow q}^{E,\text{NO}}$ and $k_{q \rightarrow p}^{E,\text{N}_2}$ are defined as: 
\begin{equation}
k_{p{\rightarrow}q}^{E,\text{NO}}=\sum_{m \in \mathcal{I}_q^{\text{N}_2}} \sum_{i \in \mathcal{I}_p^{\text{NO}}} k_{i \rightarrow m}^{E,\text{NO}}\mathcal{F}_i^{p,\text{NO}},
\label{eq:rate_p_To_q}
\end{equation}
\begin{equation}
k_{q{\rightarrow}p}^{E,\text{N}_2}=\sum_{m \in \mathcal{I}_q^{\text{N}_2}} \sum_{i \in \mathcal{I}_p^{\text{NO}}} k_{m \rightarrow i}^{E,\text{N}_2}\mathcal{F}_m^{q,\text{N}_2},
\label{eq:rate_q_To_p}
\end{equation}
where $\mathcal{F}_i^{p,\text{NO}}$ and $\mathcal{F}_m^{q,\text{N}_2}$ are, respectively,  the \emph{prescribed} internal distribution functions within  groups $p$ and $q$. Here, for simplicity, they are taken as thermal equilibrium (\emph{i.e.}, Maxwell-Boltzmann) distributions at the local translational temperature, $T$:
\begin{IEEEeqnarray}{rCl}
\mathcal{F}_i^{p,\text{NO}}&=&\frac{g_i\exp\left(-\frac{e_i}{k_BT}\right)}{\sum_{i \in \mathcal{I}_p^{\text{NO}}}g_i\exp\left(-\frac{e_i}{k_BT}\right)}=\frac{n_i}{n_p},\quad i \in \mathcal{I}_p^{\text{NO}}, \label{eq:GroupDist_NO}\\
 \mathcal{F}_m^{q,\text{N}_2}&=&\frac{g_m\exp\left(-\frac{e_m}{k_BT}\right)}{\sum_{m \in \mathcal{I}_q^{\text{N}_2}}g_m\exp\left(-\frac{e_m}{k_BT}\right)}=\frac{n_m}{n_q},\quad m \in \mathcal{I}_q^{\text{N}_2}, \label{eq:GroupDist_N2}
\end{IEEEeqnarray}
\noindent
where $e_m$ and $g_m$ denote the internal energy and degeneracy of $m$-th state, and $k_B$ stands for the Boltzmann constant. The form of $g_m$ can be found in Eq. \eqnref{eq:N2_Degeneracy} in App. \ref{app:rates}. The multiplication of both sides of Eq. \eqnref{eq:HeteroExch_p-th_q-th_Bins} by $\mathcal{F}_i^{p,\text{NO}}$ leads to: 
\begin{equation}
\frac{dn_i}{dt}=-k_{p{\rightarrow}q}^{E,\text{NO}}n_in_{\text{N}}+k_{q{\rightarrow}p}^{E,\text{N}_2}n_qn_{\text{O}}\mathcal{F}_i^{p,\text{NO}},\quad i \in \mathcal{I}_p^{\text{NO}}.
\label{eq:step-1}
\end{equation}
In Eq. \eqnref{eq:step-1}, $n_p$ is replaced by $n_i$ by invoking the local equilibrium relation \eqnref{eq:GroupDist_NO}. Similarly, the group number density $n_q$ can be transformed to the rovibrational-specific one, $n_m$, via multiplication by $\mathcal{F}_m^{q,\text{N}_2}$, which gives: 
\begin{equation}
\mathcal{F}_m^{q,\text{N}_2}\frac{dn_i}{dt}=-k_{p{\rightarrow}q}^{E,\text{NO}}\mathcal{F}_m^{q,\text{N}_2}n_in_{\text{N}}+k_{q{\rightarrow}p}^{E,\text{N}_2}\mathcal{F}_i^{p,\text{NO}}n_mn_{\text{O}},\quad i \in \mathcal{I}_p^{\text{NO}},\quad m \in \mathcal{I}_q^{\text{N}_2}.
\label{eq:step-3}
\end{equation}
From Eq. \eqnref{eq:step-3}, the grouped-reconstructed Zel'dovich reaction rate coefficients $\bar{k}_{i \rightarrow m}^{E,\text{NO}}$ and $\bar{k}_{m \rightarrow i}^{E,\text{N}_2}$ can be defined as follows:
\begin{align}
\bar{k}_{i \rightarrow m}^{E,\text{NO}} &= k_{p \rightarrow q}^{E,\text{NO}}\mathcal{F}_m^{q,\text{N}_2},\nonumber\\
\bar{k}_{m \rightarrow i}^{E,\text{N}_2} &= k_{q \rightarrow p}^{E,\text{N}_2}\mathcal{F}_i^{p,\text{NO}},\nonumber\\
\quad i \in \mathcal{I}_p^{\text{NO}} &, \quad m \in \mathcal{I}_q^{\text{N}_2}.
\label{eq:rate_Recon}
\end{align}
\noindent
This means that the grouped-reconstructed rate coefficient can be obtained from the respective grouped rate coefficient by weighting it based on the final state's contribution to its group partition function.

By spanning Eq. \eqnref{eq:step-3} over $m$ for fixed $i$, the $\mathcal{F}_m^{q,\text{N}_2}$ in the left-hand side goes to one based on the local equilibrium relation \eqnref{eq:GroupDist_N2}. Finally, the time rate of change of rovibrational-specific population, $dn_i/dt$, can be denoted using the grouped-reconstructed Zel'dovich reaction rate coefficients as follows:
\begin{equation}
\frac{dn_i}{dt}=-\bar{k}_{i \rightarrow m}^{E,\text{NO}}n_in_{\text{N}}+\bar{k}_{m \rightarrow i}^{E,\text{N}_2}n_mn_{\text{O}},\quad i \in \mathcal{I}_p^{\text{NO}}, \quad m \in \mathcal{I}_q^{\text{N}_2}.
\label{eq:step-4}
\end{equation}
Equation \eqnref{eq:step-4} allows to investigate the performance and effectiveness of reduced-order methods in reproducing the heterogeneous exchange dynamics, as the remaining excitation and dissociation channels are treated rovibrational-specifically. The corresponding results are presented and discussed in Sec. \ref{sec:all}.

\section{Results}\label{sec:results}
In this section, master equation simulations, conducted for a wide range of initial and heat-bath conditions, are analyzed and discussed in detail. Macroscopic quantities such as global rate coefficients, relaxation times, and chemistry-energy coupling parameters are calculated from the zeroth- and first-order moments of the rovibrational population distributions. Also, particular attention is devoted to investigating similarities in the population dynamics among different chemical systems and the effect of the Zel'dovich reaction on internal energy transfer at both state-specific and macroscopic levels.

In the analysis to be presented, the chemical compositions are initialized to have 50\% of diatomic target species and 50\% of atomic colliding particles, unless they are explicitly mentioned. The initial gas pressure and internal temperature are set to \SI{1000}{\pascal} and \SI{300}{\kelvin}, respectively. The above initial conditions are utilized in all simulations unless otherwise stated. As per the heat-bath temperatures, the values of \SI{2500}{\kelvin}, \SI{5000}{\kelvin}, \SI{7500}{\kelvin}, \SI{10000}{\kelvin}, \SI{15000}{\kelvin}, and \SI{20000}{\kelvin} are considered.

\subsection{Energy transfer processes in isolated systems}\label{sec:energy-iso}
In this section, energy transfer by the inelastic and homogeneous exchange in $\mathrm{N}_2$+O and NO+N systems is studied in detail by neglecting dissociation and Zel'dovich mechanisms. As a result, there is no interaction between the two chemical systems.

\begin{figure}[h]
    \centering
    \includegraphics[width=0.55\textwidth]{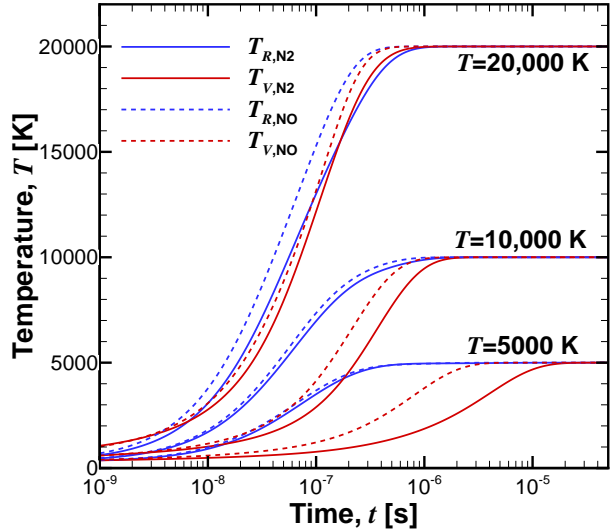}
    \caption{Temporal evolution of rotational and vibrational temperature of $\mathrm{N}_2$ and NO at different heat-bath temperatures (dissociation and Zel'dovich mechanisms excluded).}
    \label{fig:Temp_isolated}
\end{figure}

Figure \ref{fig:Temp_isolated} shows the temporal evolution of the extracted rotational ($T_R$) and vibrational ($T_V$) temperatures of $\mathrm{N}_2$ and NO at the three distinct heat-bath temperatures. The overall structure of evolution of the \emph{internal} (\emph{i.e.}, rotational and vibrational) temperatures is similar for both molecules: at the lowest temperature, vibrational relaxation is much slower compared to rotational relaxation, whereas at higher temperatures, the two processes occur at comparable time-scales. 
In the considered temperature range, both rotational and vibrational relaxation are faster for NO than $\text{N}_2$. At lower temperatures, the vibrational relaxation rates are significantly different while the rotational relaxation rates are close for the two molecules. However, as the temperature increases, the vibrational relaxation profiles for the two molecules are closer and there is larger difference in the rotational relaxation rates
The faster internal energy relaxation in NO+N is attributed to the fact that for NO the levels are more closely spaced than for $\mathrm{N}_2$, and NO collides with the lighter atomic species. Also, as opposed to the $\mathrm{N}_2$+O system, energy transfer of NO+N collisions can occur via homogeneous particle exchange in addition to the inelastic process. This contributes to the faster internal energy transfer, as observed by Panesi \emph{et al.} for the $\mathrm{N}_2$+N system \cite{PANESI_2013_BOXRVC}. 

\begin{figure}[h]
    \centering
    \includegraphics[width=0.60\textwidth]{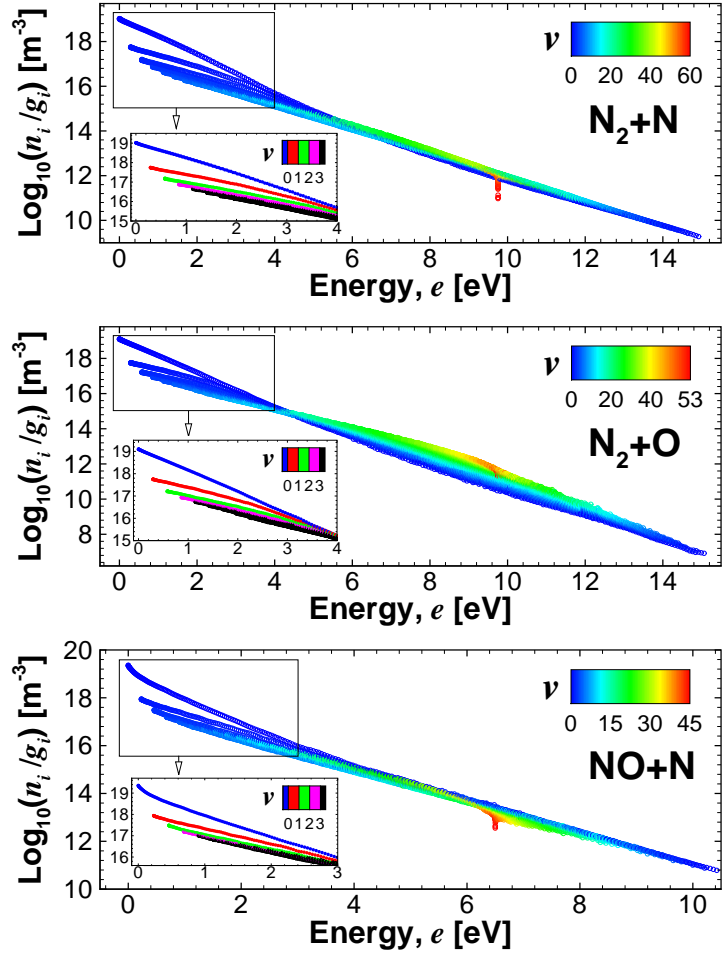}
    \caption{Rovibrational distributions in $\mathrm{N}_2$+N (top), $\mathrm{N}_2$+O (middle), and NO+N (bottom) systems at 20\% of vibrational relaxation at $T=\SI{10000}{\kelvin}$ (dissociation and Zel'dovich mechanisms excluded). The distributions are colored by the vibrational quantum number.}
    \label{fig:Pop_comp}
\end{figure}

For in-depth interpretation of the similar structure of the internal energy transfer among the different chemical systems, observed in Fig. \ref{fig:Temp_isolated}, the rovibrational distributions for the three chemical systems, $\mathrm{N}_2$+N, $\mathrm{N}_2$+O, and NO+N, are compared against each other in Fig. \ref{fig:Pop_comp}. The comparison is made at the 20\% of vibrational relaxation. It is worth mentioning that  the rovibrational rate coefficients for $\mathrm{N}_2$+N are taken from the previous work by Panesi \emph{et al.} \cite{PANESI_2013_BOXRVC}. Overall, the rovibrational distributions have similar structures even though they stem from different chemical systems. This fact is justified by the similarities between the energy transfer rate coefficients, shown in Fig. \ref{fig:InelExch_Dist}. The internal energy transfer of the diatomic species is mostly governed by the behavior of the low-lying energy states, which have a strand-like structure during the early stage of  energy transfer.\cite{PANESI_2013_BOXRVC,SAHAI_ADAPTIVE} The inset figures highlight the common existence of vibration-specific strands in the low-lying energy states and the similarity among the three chemical systems. This implies the possibility of applying the existing reduced-order method for energy transfer by Sahai \emph{et al.} \cite{SAHAI_ADAPTIVE}, originally tested on the $\mathrm{N}_2$+N interactions, to other chemical systems without a significant loss of accuracy. 

\begin{figure}[h]
    \centering
    \subfigure[]
    {
        \includegraphics[width=0.48\textwidth]{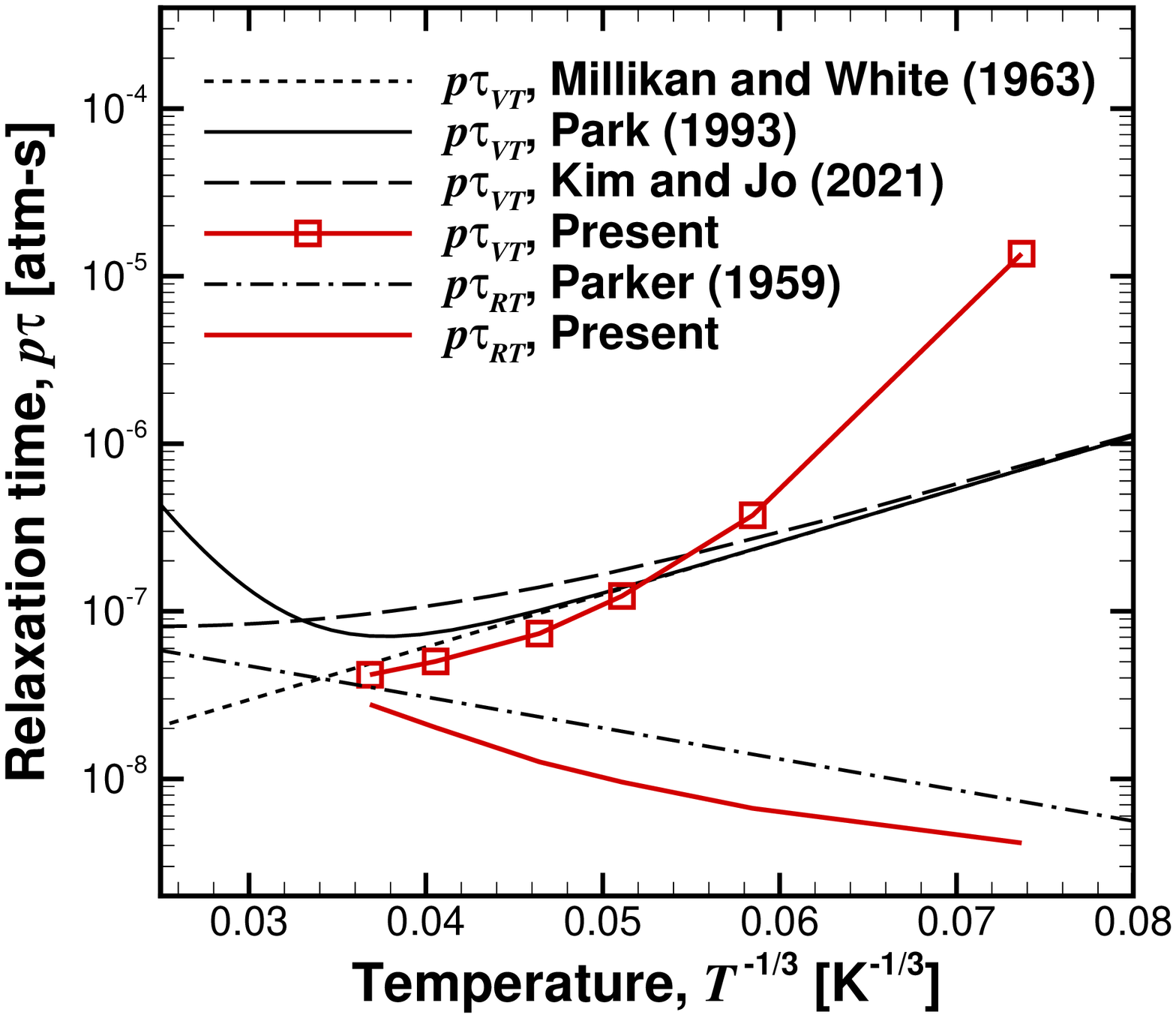}
        \label{fig:N2-O_Tau}
    }
    \subfigure[]
    {
        \includegraphics[width=0.48\textwidth]{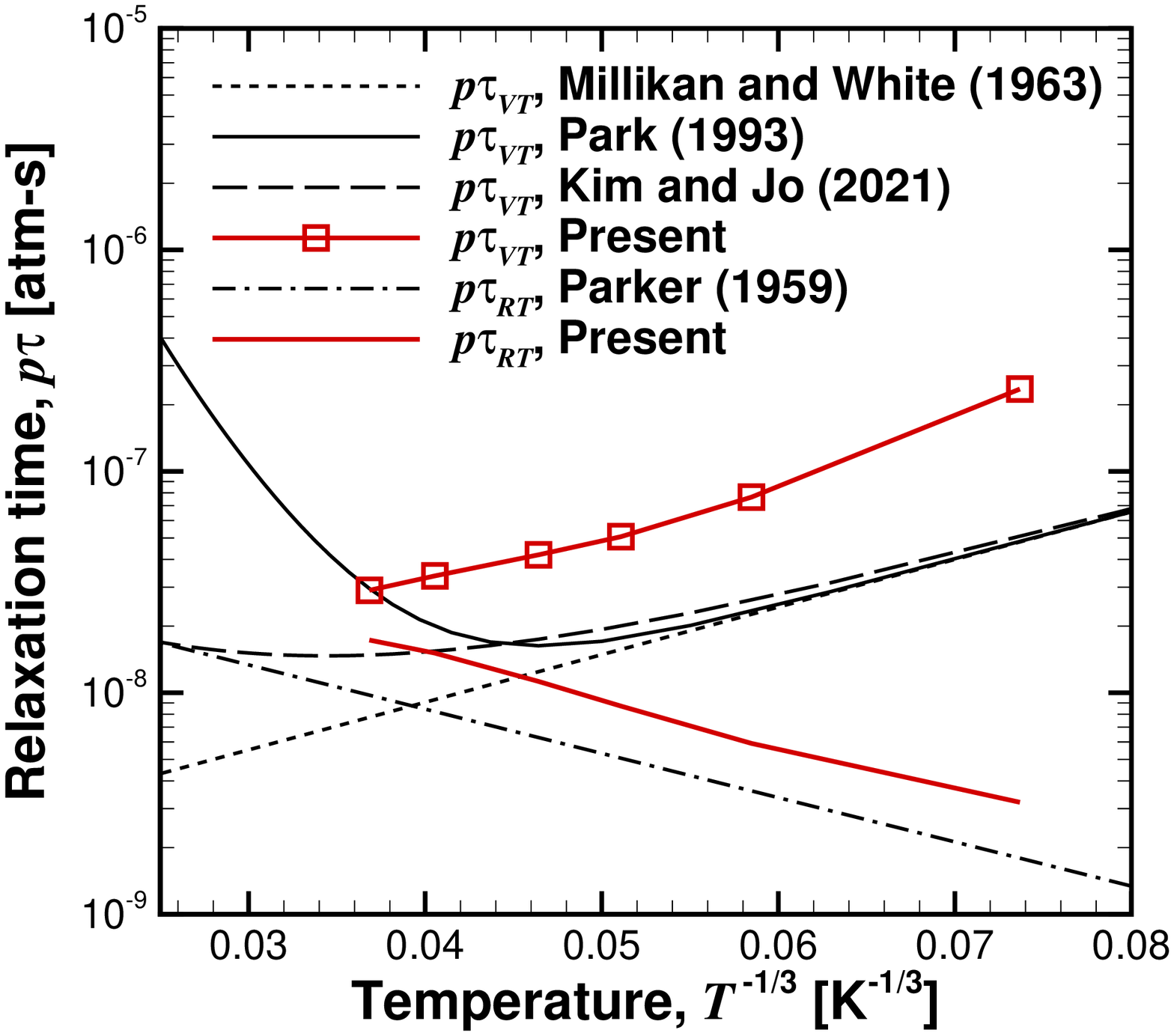}
        \label{fig:NO-N_Tau}
    }
  \caption{Comparison of predicted rotational and vibrational relaxation times with existing theoretical models \cite{MW_1963,Park_1993_Earth,Kim_2021,Parker_1959}: (a) $\mathrm{N}_2$+O, (b) NO+N (dissociation and Zel'dovich mechanisms excluded).}
    \label{fig:Tau_woHetExch}
\end{figure}

One of the prevalent ways of describing energy transfer in hypersonic environments is via the Landau-Teller model,\cite{LT_1936} which requires specifying a relaxation time $\tau$ for the kinetic process being considered. The value of $\tau$ can affect the behavior of thermal energy relaxation in the non-equilibrium region of hypersonic shock layers. Figure \ref{fig:Tau_woHetExch} comparers the rotational ($p \, \tau_{RT}$) and vibrational ($p \, \tau_{VT}$) relaxation times of the $\mathrm{N}_2$+O and NO+N systems as a function of the kinetic temperature. Here $p$ denotes the partial pressure of the colliding atomic species. In the present study, the relaxation time is evaluated based on the \emph{e-folding} method. To the best of the authors' knowledge, there is no available experimental data for rotational and vibrational relaxation times for both chemical systems. Thus, the comparison is performed against existing theoretical models.\cite{MW_1963,Park_1993_Earth,Kim_2021,Parker_1959} 

For both chemical systems, the correlation-based model by Millikan and White, \cite{MW_1963} later modified by Park for high-temperature effects, \cite{Park_1993_Earth} shows a strong departure from the values computed here. For $\mathrm{N}_2$+O, the temperature dependence of $p \, \tau_{VT}$ from the present result is different from those of the previous studies,\cite{MW_1963,Park_1993_Earth,Kim_2021} especially below \SI{5000}{\kelvin}. For NO+N, the temperature dependence of the present result is similar to those of existing studies, \cite{MW_1963,Park_1993_Earth,Kim_2021} although the absolute value of the present $p \, \tau_{VT}$ is larger by a factor of 2.5. As per rotational relaxation, the present results are in disagreement with the Parker model \cite{Parker_1959} by around a factor of 2 for both chemical systems. In the Parker model, the rotational relaxation time was derived by classical mechanics along using the rigid-rotor model that cannot account for the effect of a permanent dipole in diatom-atom systems. \cite{Parker_1959,Jo_2021} Consistent with the temperature evolution of Fig. \ref{fig:Temp_isolated}, the rotational and vibrational relaxation times of both chemical systems converge to a common asymptote as heat-bath temperature increases. The asymptote of the rotational and vibrational relaxation times at the high-temperature range was also observed in the previous studies of the $\mathrm{N}_2$+N system \cite{PANESI_2013_BOXRVC,Kim_Boyd_Cphys_2013}. In those references, the asymptote was justified by the presence of the homogeneous exchange reaction. However, it is also observed here for the $\mathrm{N}_2$+O system, in which a homogeneous exchange process does not occur. 

\subsection{Dissociation/recombination processes in isolated systems}\label{sec:all-iso}
In this section, dissociation and recombination processes in $\mathrm{N}_2$+O and NO+N systems are investigated by disregarding the Zel'dovich mechanism. Hence, the two systems evolve independently as in the previous section.

\begin{figure}[h]
    \centering
    \includegraphics[width=0.58\textwidth]{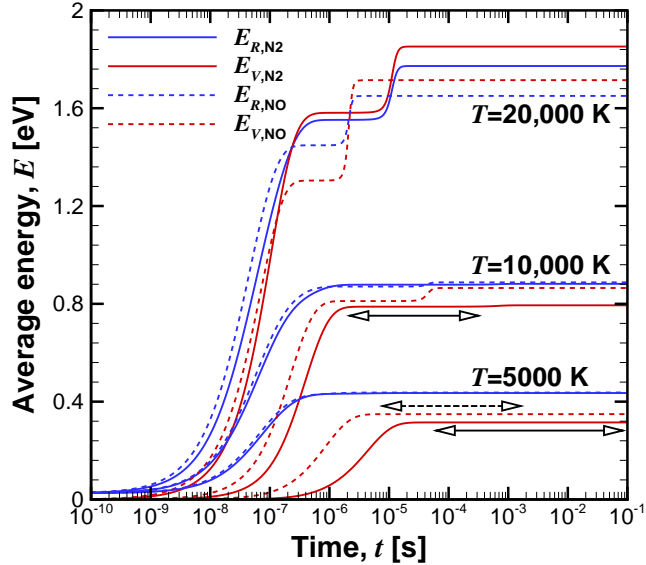}
    \caption{Average rotational and vibrational energy distributions of $\text{N}_2$ and NO at the three different kinetic temperatures (Zel'dovich mechanism excluded). The arrows indicate the molecular QSS periods in cases that the plateau regime is not explicitly shown. The solid arrow indicates the $\text{N}_2$ QSS, while the dashed one denotes the NO QSS.}
    \label{fig:Energy_isolated}
\end{figure}

Figure \ref{fig:Energy_isolated} shows temporal evolution of the average rotational ($E_R$) and vibrational ($E_V$) energy of $\text{N}_2$ and NO, at three different kinetic temperatures, $T$. In the present study, the average rotational and vibrational energy of $\text{N}_2$ are correspondingly defined as
\begin{equation}
E_{R,\text{N}_2}=\frac{\sum_m^{\text{N}_2}e_r\left(m\right)n_m}{\sum_m^{\text{N}_2}n_m},
\label{eq:Average_Rotational_Energy_N2}
\end{equation}
\begin{equation}
E_{V,\text{N}_2}=\frac{\sum_m^{\text{N}_2}e_v\left(m\right)n_m}{\sum_m^{\text{N}_2}n_m},
\label{eq:Average_Vibrational_Energy_N2}
\end{equation}
\noindent
where $e_r\left(m\right)$ and $e_v\left(m\right)$ are the rotational and vibrational energy of the given rovibrational state $m$, respectively. The plateau of the energy curves indicate the QSS periods of each species. As discussed in the following paragraphs, most of the dissociation occurs during the QSS periods. The following evolution from the plateau to the equilibrium state is governed by the recombination processes. From the figure, it is noted that the time scale of NO recombination is faster than that of $\text{N}_2$ over $T=\SI{10000}{\kelvin}$. Similar to the energy transfer processes, shown in Fig. \ref{fig:Temp_isolated}, the energy evolution trends of $\text{N}_2$ and NO are close to each other, while NO represents the faster onset of the QSS period than $\text{N}_2$ due to the lower dissociation limit and the faster internal energy transfers within NO. Furthermore, since $\text{N}_2$ has a lager bonding energy than NO, the QSS period of $\text{N}_2$ is longer than that of NO.

\begin{figure}[h]
    \centering
    \includegraphics[width=0.75\textwidth]{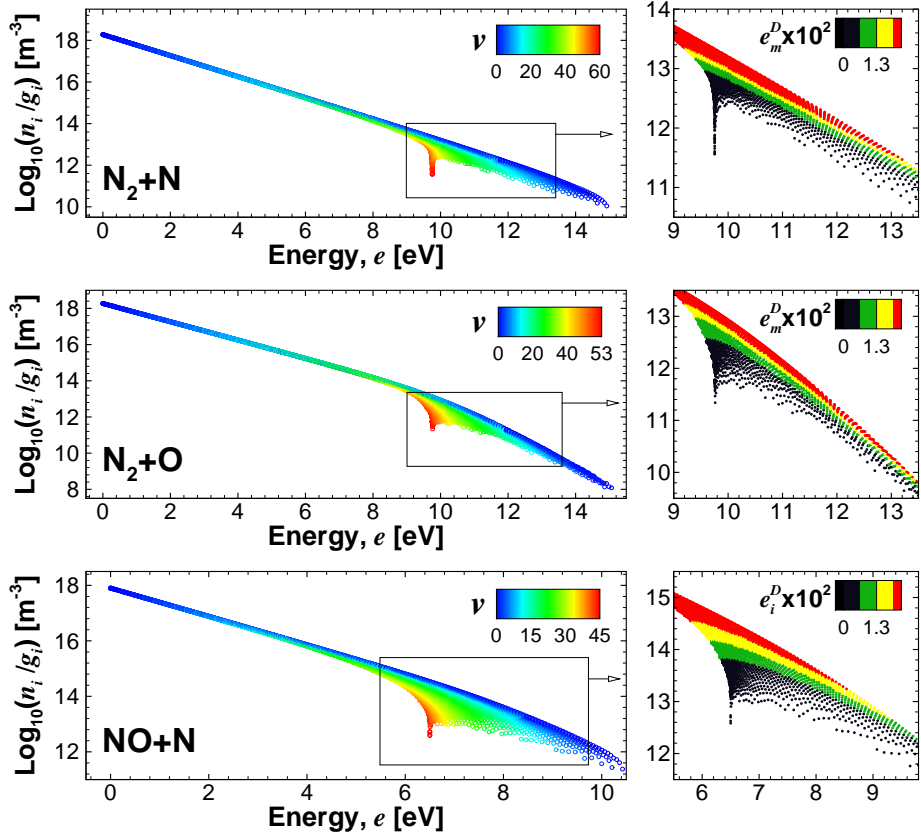}
    \caption{Rovibrational distributions in $\text{N}_2$+N (top), $\text{N}_2$+O (middle), and NO+N (bottom) systems at midst of QSS region at $T=\SI{10000}{\kelvin}$ (Zel'dovich mechanism excluded). The distributions are colored by the vibrational quantum number $v$ and the energy deficit $e^D$.}
    \label{fig:Pop_comp_QSS}
\end{figure}

The comparison among the internal state population distributions from the different chemical systems, shown in Fig. \ref{fig:Pop_comp}, is re-examined here for the dissociation and recombination processes. In Fig. \ref{fig:Pop_comp_QSS}, a comparison of the internal energy state population distributions is presented for the $\text{N}_2$+N, $\text{N}_2$+O, and NO+N systems during the QSS period. The main figures (\emph{i.e.}, those in the left-side) are colored by the vibrational quantum number $v$, while the sub-figures (\emph{i.e.}, those in the right-side) are colored by the state-specific energy deficit from the centrifugal barrier. In the present study, the energy deficit of $\text{N}_2$ and NO are correspondingly defined as \cite{Venturi2020}
\begin{equation}
e_m^D=V_{\text{max}}^{\text{N}_2}\left(J\right)-e_m,
\label{eq:Energy_Deficit_N2}
\end{equation}
\begin{equation}
e_i^D=V_{\text{max}}^{\text{NO}}\left(J\right)-e_i,
\label{eq:Energy_Deficit_N2}
\end{equation}
\noindent
where $V_{\text{max}}^{\text{N}_2}\left(J\right)$ and $V_{\text{max}}^{\text{NO}}\left(J\right)$ denote the maximum values of the effective diatomic potential at the given rotational state $J$ of $\text{N}_2$ and NO, respectively. As observed from the $v$ contours, the vibrational states, which are in the high-lying energy region around the dissociation limit, are not in equilibrium with each other. This indicates that the collisional dissociation processes are not governed by a vibration-specific dynamics. On the other hand, the subfigures clearly show that the equilibrium distributions of the high-lying energy states is highly correlated with the levels' energy deficit. This is in line with the previous study by Venturi \emph{et al.} \cite{Venturi2020} regarding the $\text{O}_2$+O system, and it is another evidence of the existence of similarity among different chemical systems, especially for the collisional dissociation process.

In multi-dimensional hypersonic flow-field simulations, one of the most popular strategies to model the collisional chemical production rates $\dot{\omega}$ is to employ the macroscopic QSS rate coefficient by assuming that the participating species are in the QSS region\cite{Park_book}. The rovibrational-specific master equation analysis performed in the present study allows us to evaluate the actual amount of chemical reactions taking place during the QSS periods of the participating species, and to verify the QSS assumption for the chemical reaction modeling. In this section, it is performed for the collisional dissociation processes in the isolated chemical systems, while the complete chemical system including the Zel'dovich mechanism is investigated in Sec. \ref{sec:all}. 

\begin{figure}[h]
    \centering
    \subfigure[]
    {
        \includegraphics[width=0.48\textwidth]{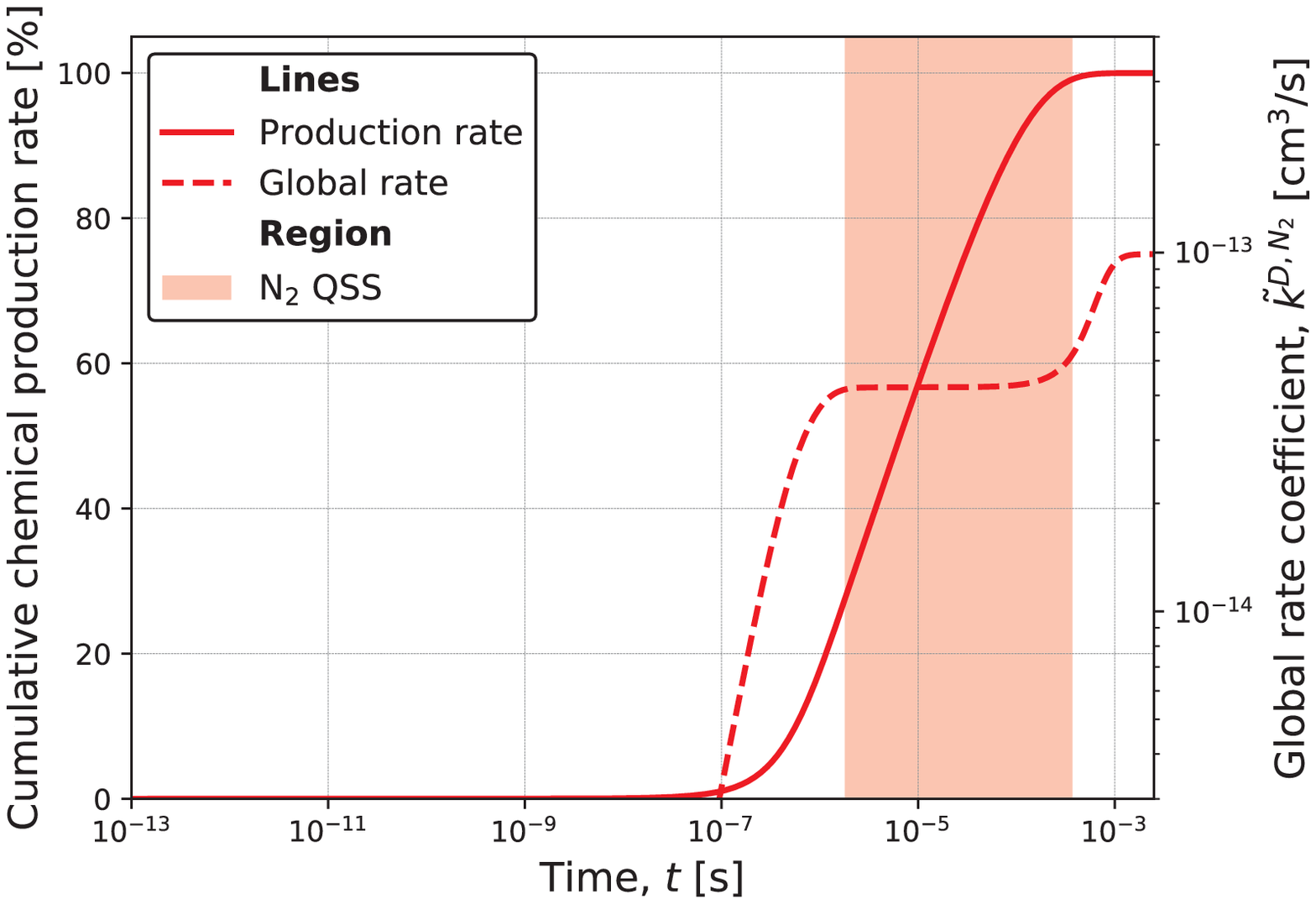}
        \label{fig:N2-Diss_Region}
    }
    \subfigure[]
    {
        \includegraphics[width=0.48\textwidth]{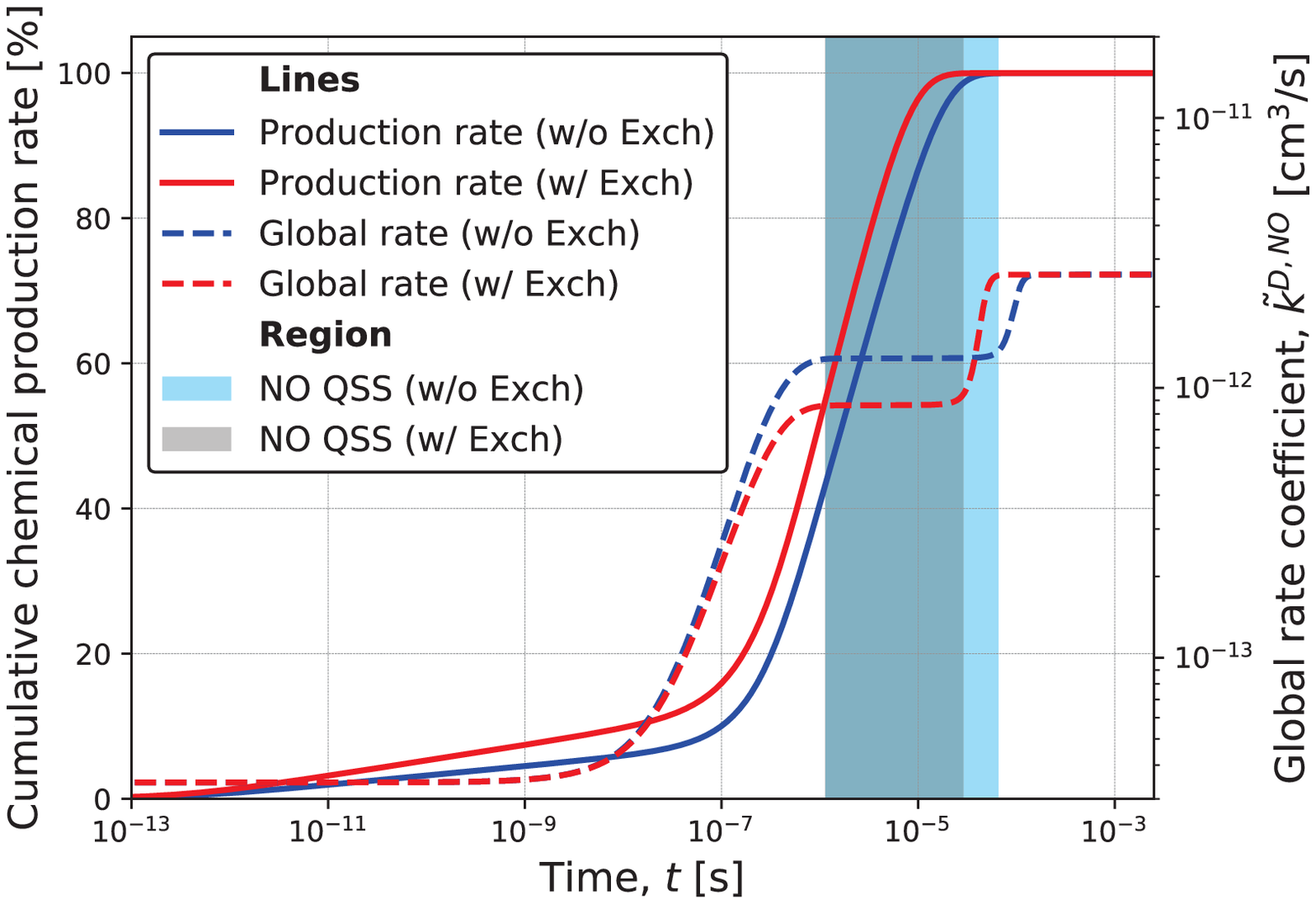}
        \label{fig:NO-Diss_Region}
    }
  \caption{Time-cumulative chemical production and global dissociation rate distributions at $T=\SI{10000}{\kelvin}$ (Zel'dovich mechanism excluded). The solid lines indicate the chemical production rates, while the dashed lines represent the global dissociation rate coefficients. The shaded regions indicate the QSS periods of diatomic species. (a) $\text{N}_2$+O, (b) NO+N.}
    \label{fig:Diss_Isolated}
\end{figure}

Figure \ref{fig:Diss_Isolated} shows the time-cumulative chemical production rate distributions for the dissociation and recombination processes at $T=\SI{10000}{\kelvin}$, together with the global dissociation rate coefficients. The time-cumulative quantity was defined by summing up and normalizing the $\dot{\omega}^D$ at given time-step with the total amount of chemical production rate due to dissociation and recombination. It needs to be noted that the initial species mole fractions for the heat-bath simulations are different from each other to investigate the two different isolated chemical systems. For the dissociation of $\text{N}_2$ by $\text{N}_2$+O collisions, about 70\% of the dissociation takes place during the QSS period of $\text{N}_2$ as shown in Fig. \ref{fig:N2-Diss_Region}. On the other hand, about 50\% of NO dissociation occurs within the QSS period.
For NO+N, three different reaction channels exist for the dissociation due to the possibility of the homogeneous and heterogeneous exchanges prior to the dissociation. This aspect degrades the validity of QSS assumption for NO dissociation, as shown in Fig. \ref{fig:NO-Diss_Region}. The contribution of the exchanged pairs to dissociation leads to about 10\% difference in the amount of dissociation taking place  in the QSS period. It needs to be noted that the levels of influence from the heterogeneous and the homogeneous exchanged pairs are close to each other. The influence of the exchange pair on dissociation of NO shown in Fig. \ref{fig:NO-Diss_Region} is consistently observed at $T=\SI{5000}{\kelvin}$ and \SI{20000}{\kelvin} as well (See Fig. S1 in Supplementary Material). Additionally, faster vibrational energy transfer in NO, as observed in Fig. \ref{fig:Temp_isolated}, further reduces the applicability of the QSS assumption for NO dissociation.
The faster vibrational ladder-climbing of NO contributes to the faster dissociation since the dissociation-recombination processes are mostly controlled by the high-lying rovibrational states.

\begin{figure}[h]
    \centering
    \includegraphics[width=0.55\textwidth]{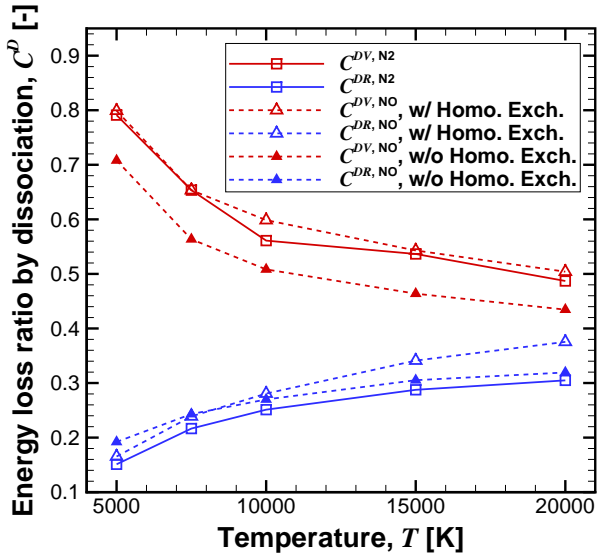}
    \caption{Rotational and vibrational energy loss ratios due to the dissociation as a function of kinetic temperatures (Zel'dovich mechanism excluded).}
    \label{fig:Diss-EnergyCoupling}
\end{figure}

In Fig. \ref{fig:Diss-EnergyCoupling}, the internal energy loss ratio due to the dissociation, $C^D$, is presented as a function of $T$. The definition of $C^D$ is taken from the previous study by Panesi \emph{et al.} \cite{PANESI_2013_BOXRVC}, and the quantity is defined at the mid-point of the species QSS periods. 
The vibrational energy loss ratios are very similar between $\text{N}_2$ and NO, provided that the NO+N system has both the inelastic and homogeneous exchange processes. On the other hand, the NO dissociation has a larger portion of rotational energy loss than $\text{N}_2$, especially in high $T$ range. A part of this trend is attributed to the existence of homogeneous exchange reaction in the NO+N system. This fact accelerates the thermalization among the vibrational states, resulting in more populated high-lying $J$ levels (See Fig. S2 in Supplementary Material). Without the homogeneous exchange, the rotational energy loss ratio of NO becomes similar to that of $\text{N}_2$, whereas the vibrational energy loss shifts to lower values with a certain amount of offset due to the less population of high-lying $v$ states. As $T$ increases, the rotational and vibrational energy loss ratios move in the opposite direction. This is because the contribution from the internal states characterized by lower $v$ and higher $J$ becomes dominant as $T$ increases. It is interesting to note that, for both $\text{N}_2$ and NO, the vibrational energy loss ratios $C^{DV}$ are rather different from the previously estimated value of 0.3 using SSH theory \cite{Sharma_1988} in the present considered temperature range.

\subsection{Zel'dovich mechanism}\label{sec:all}
In this section, the results of master equation analyses are presented by considering the complete set of chemical-kinetic processes, Eq. (\ref{eq:NO_Inel-HomoExch}) to Eq. (\ref{eq:N2O_HetExch}). This allows us to investigate the role of collisional heterogeneous exchange process, Eq. (\ref{eq:N2O_HetExch}), through the forward and backward Zel'dovich mechanisms in the internal energy transfers of the $\text{N}_2$+O and NO+N systems, and to characterize the NO formation and extinction. For this purpose, the set of master equations from Eq. (\ref{eq:NO_Pop}) to Eq. (\ref{eq:N_Pop}) are integrated, and none of the individual components are set to zero in this case. In total, approximately an order of $10^8$ kinetic processes is considered for solving the set of master equations in the considered temperature range.

\begin{figure}[h]
    \centering
    \includegraphics[width=0.55\textwidth]{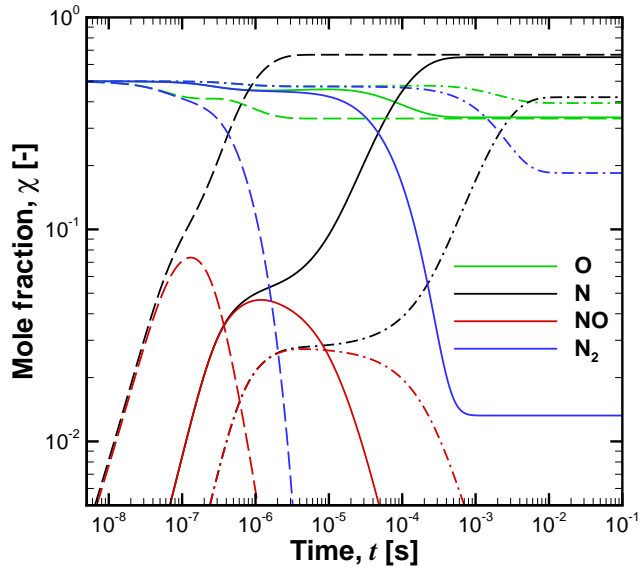}
    \caption{Species mole fraction distributions at $T=\SI{7500}{\kelvin}$ (dashed-dot lines), \SI{10000}{\kelvin} (solid lines), and \SI{20000}{\kelvin} (dashed lines). The species initial mole fractions are set to $\chi_{\text{N}_2}=\chi_{\text{O}}$=0.5.}
    \label{fig:HetExch_MoleFrac}
\end{figure}

\begin{figure}[h]
    \centering
    \subfigure[]
    {
        \includegraphics[width=0.48\textwidth]{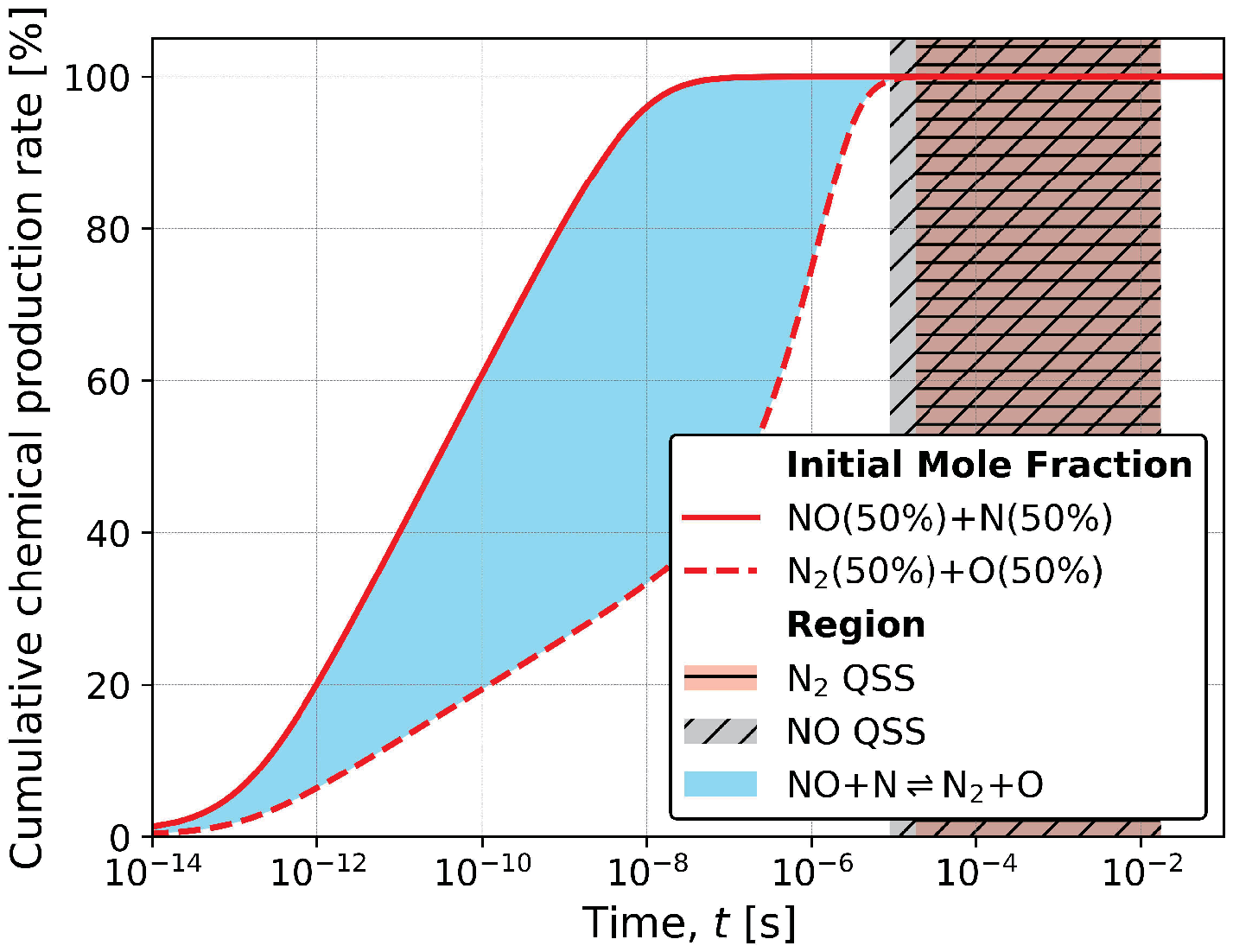}
        \label{fig:Exch_Region_5000K}
    }
    \subfigure[]
    {
        \includegraphics[width=0.48\textwidth]{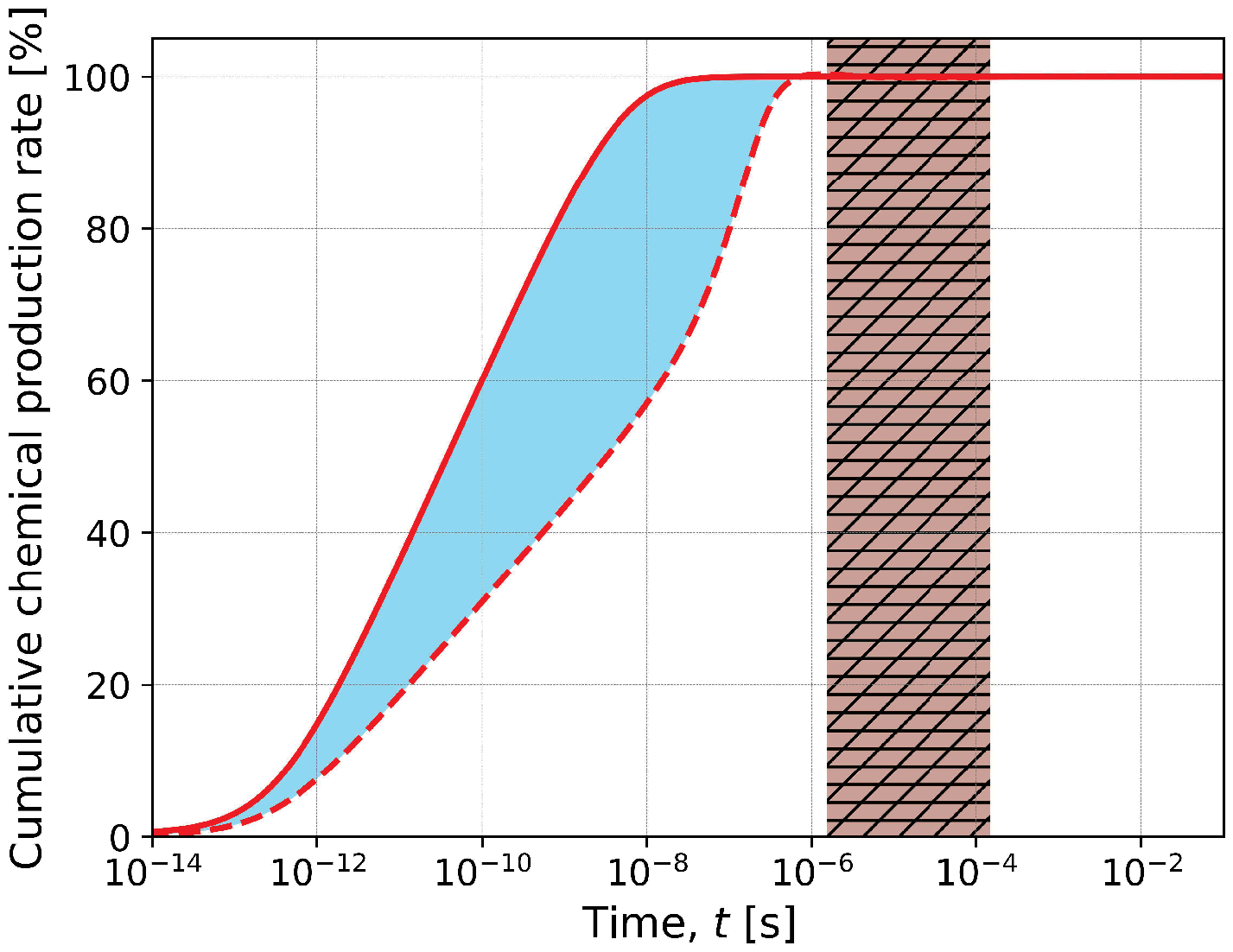}
        \label{fig:Exch_Region_10000K}
    }
    \subfigure[]
    {
        \includegraphics[width=0.48\textwidth]{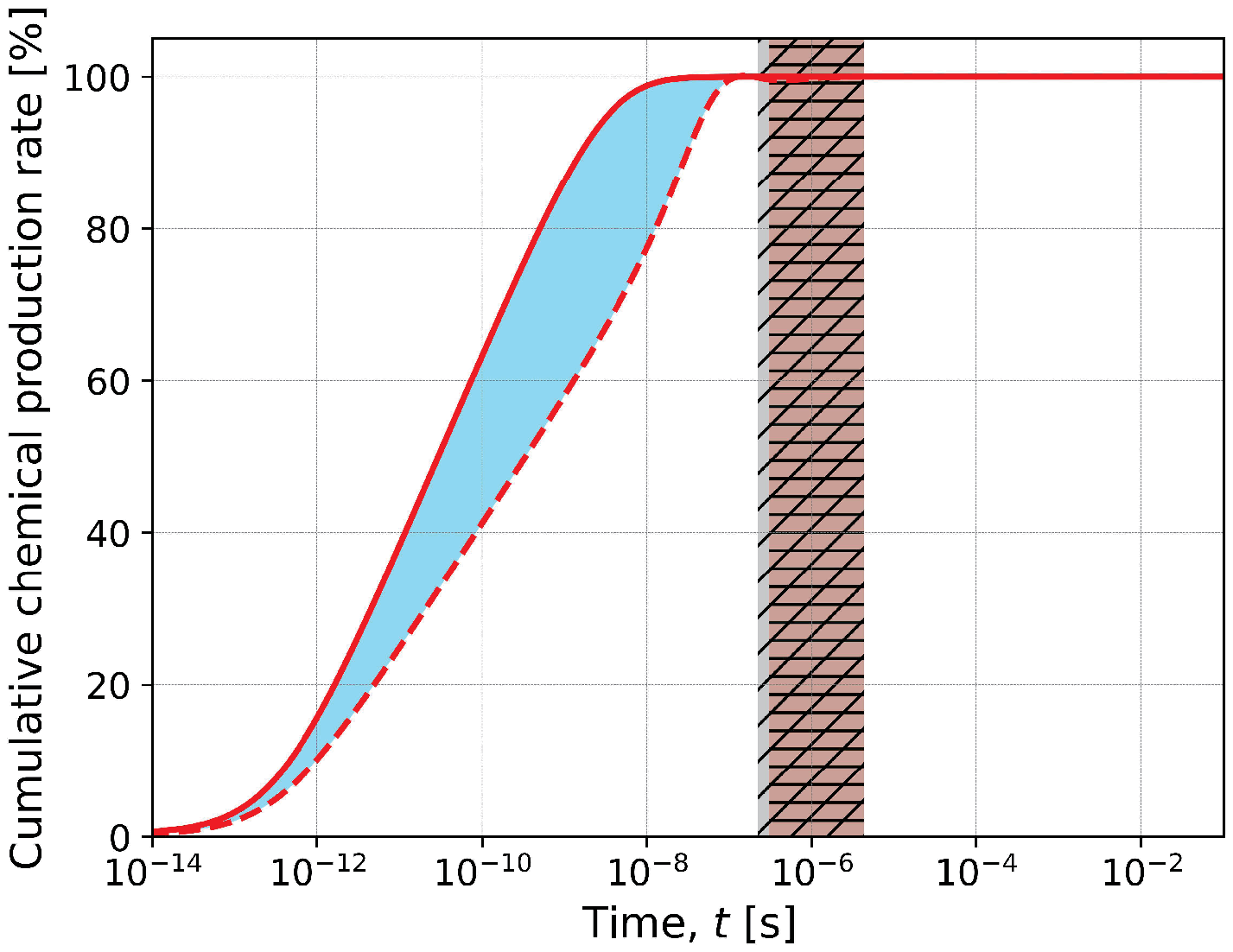}
        \label{fig:Exch_Region_20000K}
    }
  \caption{Time-cumulative chemical production rate distributions for the heterogeneous exchange process NO+N $\leftrightarrow$ $\text{N}_2$+O. The magenta and gray shaded regions indicate the QSS periods of $\text{N}_2$ and NO, respectively. The cyan shaded areas represent the region where the chemical reaction occurs: (a) $T=\SI{5000}{\kelvin}$, (b) $T=\SI{10000}{\kelvin}$, (c) $T=\SI{20000}{\kelvin}$. The red lines indicate the two different limiting cases of initial mole fractions. The figures share the plot legend.}
    \label{fig:Exch_Region_All}
\end{figure}

Figure \ref{fig:HetExch_MoleFrac} shows temporal evolution of species mole fractions $\chi$ at the three different kinetic temperatures $T$. As $T$ increases, the level of NO formation increases due to the larger contribution from the Zel'dovich mechanism. In the early stages, nearly until the peaks of $\chi_{\text{NO}}$, the mole fraction profiles of the diatomic species are aligned together with their atomic collision partners because the collisional heterogeneous exchange actively occurs prior to the onset of the molecular QSS (See Fig. \ref{fig:Exch_Region_All} for more details). Once most of the heterogeneous exchange reaction occurs, the mole fractions of each collision pair are split into two branches. This fact indicates that the dissociation starts to control the overall chemical reactions. It is interesting to note that the formation of NO is governed by the heterogeneous exchange from $\text{N}_2$, whereas the extinction of NO is mostly controlled by the dissociation of NO rather than the exchange kinetics into $\text{N}_2$.

\begin{figure}[h]
    \centering
    \subfigure[]
    {
        \includegraphics[width=0.8\textwidth]{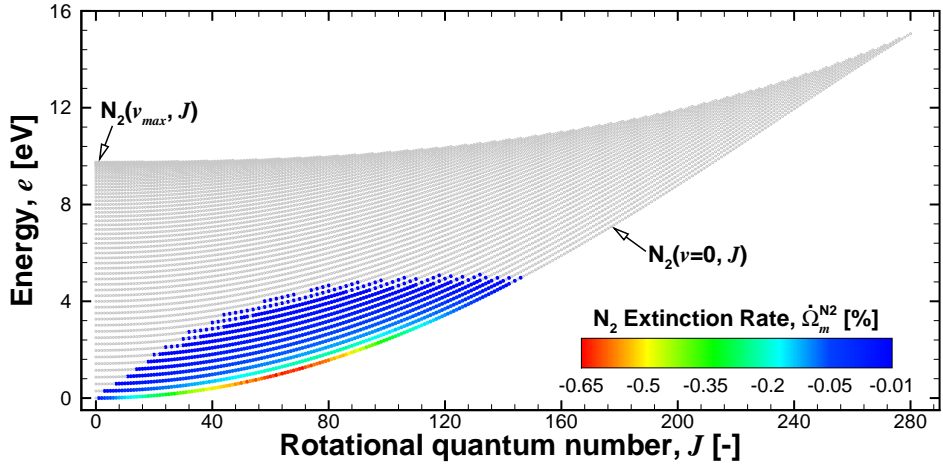}
        \label{fig:HetExch_N2Source}
    }
    \subfigure[]
    {
        \includegraphics[width=0.8\textwidth]{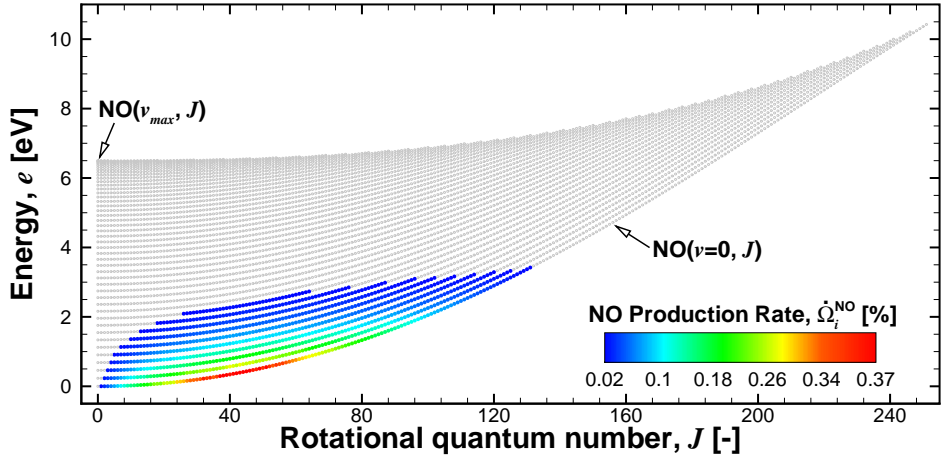}
        \label{fig:HetExch_NOSource}
    }
  \caption{Percentage distributions of the heterogeneous exchange production rates at $t$=2$\times$10$^{-7}$ s of $T=\SI{10000}{\kelvin}$ in Fig. \ref{fig:HetExch_MoleFrac}. The empty gray dots indicate that the production rate is below or above a cutoff value. (a) $\text{N}_2$ extinction rate with a cutoff value above -0.01\%, (b) NO production rate with a cutoff value below 0.02\%.}
    \label{fig:HetExch}
\end{figure}

The mechanism of NO formation is worth to be explored in detail since NO is a strongly radiating species in a hypersonic flow regime, and it has an important role in high-temperature combustion and related air pollution study. Figure \ref{fig:HetExch} provides the rovibrationally resolved production rates of connected to $\text{N}_2$ extinction. The reported quantity was computed from the result of master equation study at $T=\SI{10000}{\kelvin}$ and $t$=2$\times$10$^{-7}$ s of Fig. \ref{fig:HetExch_MoleFrac}. The state-specific $\text{N}_2$ extinction and NO formation rates are correspondingly defined by summing up and normalizing Eq. (\ref{eq:Source_HeteroExch}) as follows
\begin{equation}
\dot{\Omega}_m^{\text{N}_2}=\frac{100 \times \sum_i^{\text{NO}}\dot{\omega}_{i,m}^{E,\text{NO}}}{\left| \sum_m^{\text{N}_2}\sum_i^{\text{NO}}\dot{\omega}_{i,m}^{E,\text{NO}}\right|},
\label{eq:N2_Extinction}
\end{equation}
\begin{equation}
\dot{\Omega}_i^{\text{NO}}=-\frac{100 \times \sum_m^{\text{N}_2}\dot{\omega}_{i,m}^{E,\text{NO}}}{\left| \sum_m^{\text{N}_2}\sum_i^{\text{NO}}\dot{\omega}_{i,m}^{E,\text{NO}}\right|}.
\label{eq:NO_Formation}
\end{equation}
\noindent
Figure \ref{fig:HetExch_N2Source} indicates that the low-lying states of $\text{N}_2$ around $v$=0 and $J$=65 have major contributions for the conversion into NO during the NO formation phase. Similarly, the NO production occurs with the predominant influence from the low-lying states near $v$=0 and $J$=45 as shown in Fig. \ref{fig:HetExch_NOSource}. After the conversion of $\text{N}_2$ into the low-lying states of NO, the inelastic and homogeneous exchange by NO+N collision govern the ladder-climbing excitation process to the high-lying energy levels.

Figure \ref{fig:Exch_Region_All} shows the time-cumulative chemical production rate distributions by the Zel'dovich mechanism at $T$=\SI{5000}{\kelvin}, \SI{10000}{\kelvin}, and \SI{20000}{\kelvin}. It is important to note that the red lines indicate the limiting boundaries of the variation of chemical reactions. 
At a given temperature and pressure, the master equation results for different initial sets of mole fractions fall within the cyan region in Fig.~\ref{fig:Exch_Region_All}, provided that the mole fraction ratio of the target molecules and colliding atoms is one.
As shown in the figures, almost all of the Zel'dovich processes occur outside the molecular QSS periods of the considered temperature range. This implies that the QSS assumption is an inadequate approximation for the Zel'dovich reaction $\text{N}_2$+O $\leftrightarrow$ NO+N. In addition, the feasibility of the QSS assumption is further deteriorates as more NO exists at the initial condition.

\begin{figure}[h]
    \centering
    \subfigure[]
    {
        \includegraphics[width=0.48\textwidth]{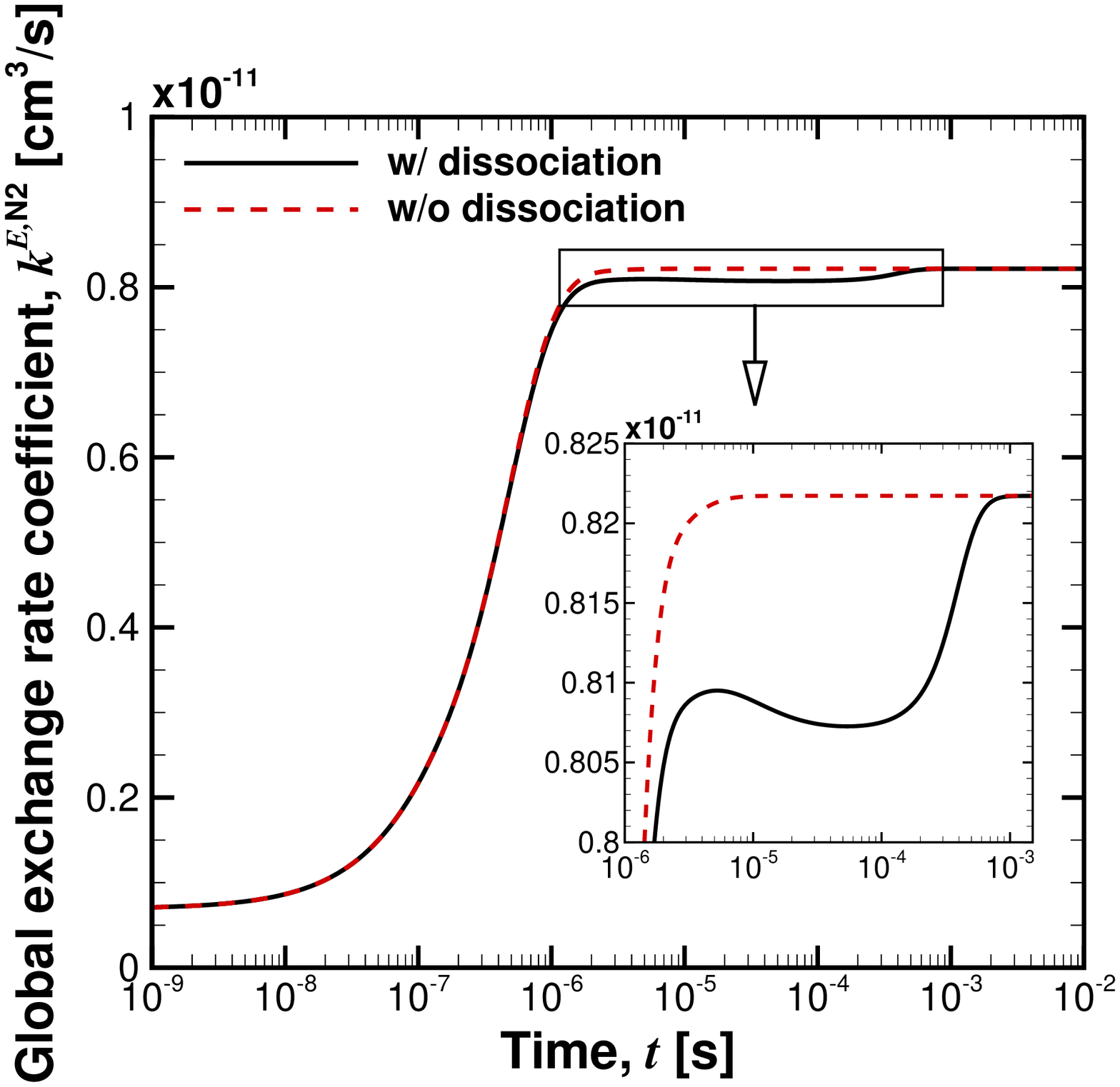}
        \label{fig:GlobalRate_T10000K}
    }
    \subfigure[]
    {
        \includegraphics[width=0.478\textwidth]{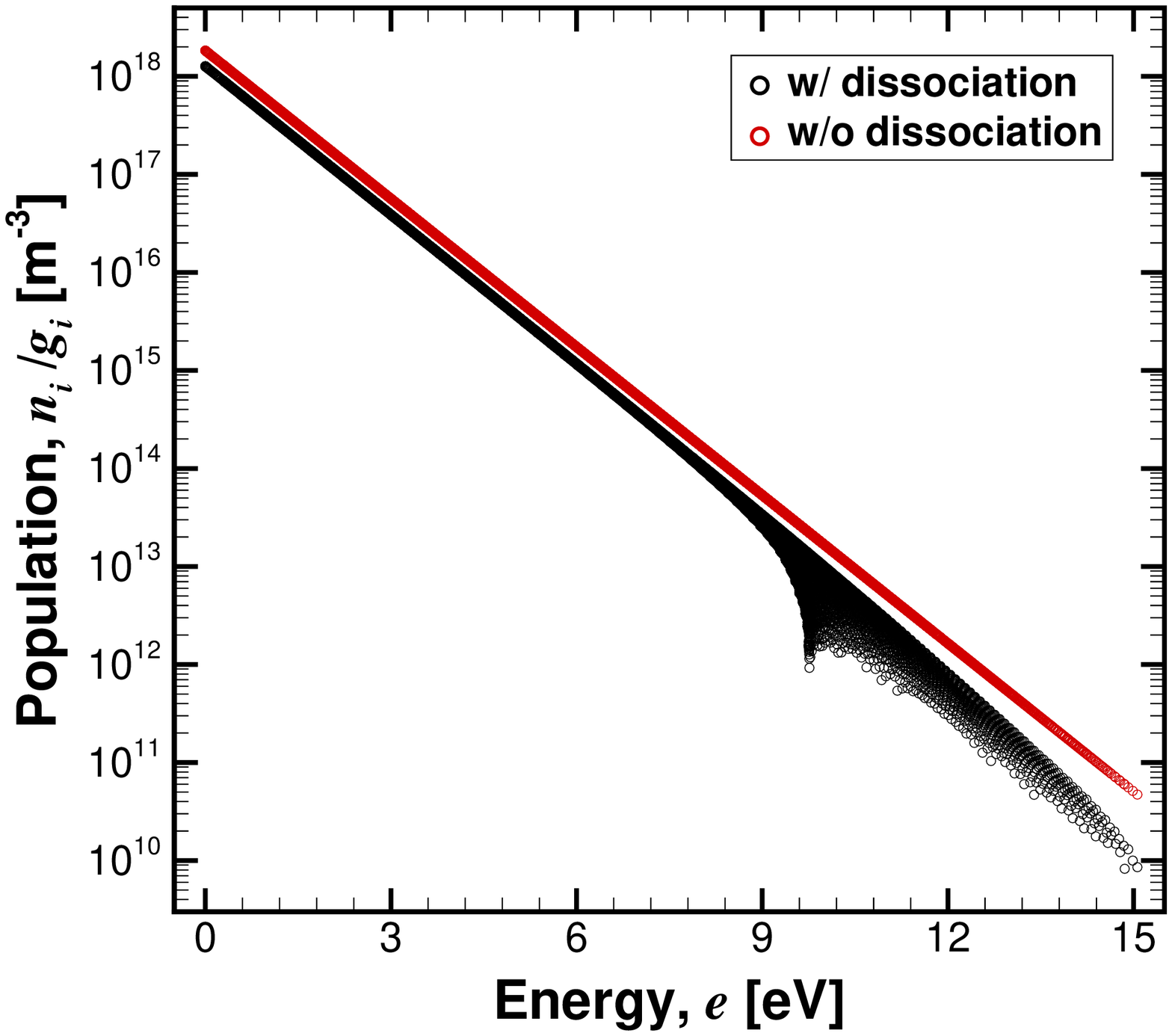}
        \label{fig:N2-Pop_T10000K}
    }
  \caption{Comparisons of (a) the global rate coefficient for $\text{N}_2$+O $\rightarrow$ NO+N, and (b) the corresponding rovibrational distributions of $\text{N}_2$ at $t$=1.7$\times$10$^{-5}$ s at $T$=\SI{10000}{\kelvin} with and without dissociation kinetics. The initial mole fraction is set to $\chi_{\text{N}_2}$=$\chi_{\text{O}}$=0.5.}
    \label{fig:NoQSS_withoutDiss}
\end{figure}

Since the majority of the Zel'dovich reaction occurs outside of the molecular QSS regime, this process requires further investigations to better understand which processes contribute to QSS. To this aim, we analyze the contribution of dissociation to the Zel'dovich mechanism. This is based on the fact that the previous studies \cite{PANESI_2013_BOXRVC,Kim_Boyd_Cphys_2013} have revealed that the dissociation predominantly promotes to the formation of QSS. Figure \ref{fig:NoQSS_withoutDiss} compares the global rate coefficient for $\text{N}_2$+O $\rightarrow$ NO+N and the corresponding rovibrational distributions with and without dissociation. The definition of $\tilde{k}^{E,\text{N}_2}$ can be found in Eq. \eqnref{eq:GlobalHetExchRate_N2_To_NO}. As shown in Fig. \ref{fig:GlobalRate_T10000K}, the global rate coefficient without the dissociation does not have the plateau region (\emph{i.e.}, from $t$=2$\times$10$^{-6}$ s to $t$=2$\times$10$^{-4}$ s for the black solid line). This means that the system does not reach the QSS period if the dissociation process is not considered, as supported by the rovibrational distribution shown in Fig. \ref{fig:N2-Pop_T10000K}. At the same time instant, the case without dissociation already reaches the thermal equilibrium, whereas the rovibrational states are aligned as the QSS distribution in the case with dissociation. The result in Fig. \ref{fig:NoQSS_withoutDiss} is a clear evidence of the breakdown of the QSS assumption for modeling the chemical reaction by the Zel'dovich mechanism \eqnref{eq:NO_Zel'dovich}, and that the QSS period is primarily controlled by the dissociation process. Results for a similar investigation for $T$=\SI{5000}{\kelvin} and \SI{20000}{\kelvin} can be found in Figs. S3 and S4 in Supplementary Material.

\begin{figure}[h]
    \centering
    \includegraphics[width=0.65\textwidth]{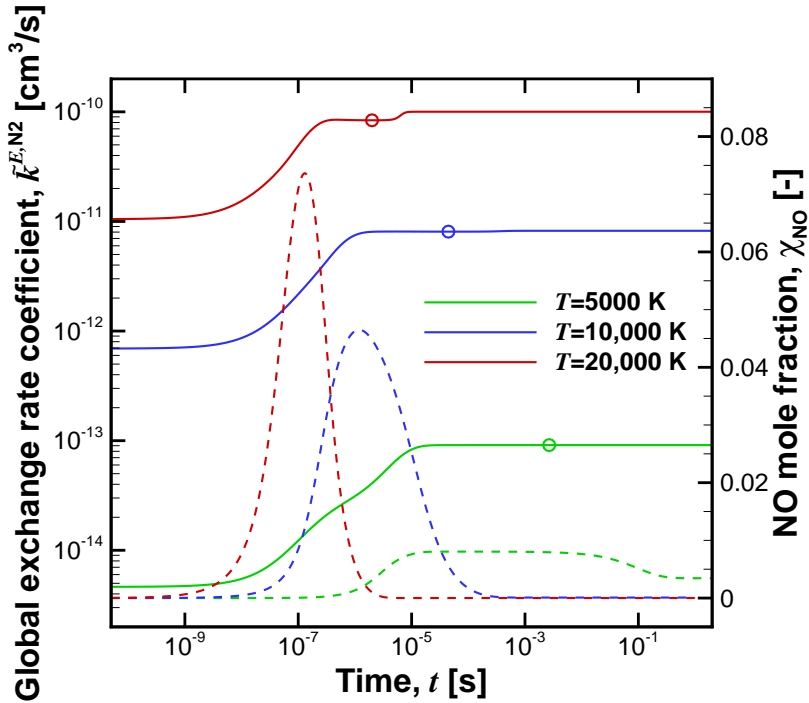}
    \caption{Time evolution of the global rate coefficient for $\text{N}_2$+O $\rightarrow$ NO+N (solid lines), and corresponding NO mole fractions (dashed lines) at $T=\SI{5000}{\kelvin}$, \SI{10000}{\kelvin}, and \SI{20000}{\kelvin}. The empty dots point out the middle of the $\text{N}_2$ QSS region. The initial mole fraction was set to $\chi_{\text{N}_2}$=$\chi_{\text{O}}$=0.5.}
    \label{fig:GlobalN2O_NOMole}
\end{figure}

As observed in Figs. \ref{fig:Exch_Region_All} and \ref{fig:NoQSS_withoutDiss}, the QSS approximation breakdowns for the Zel'dovich mechanism. This requires further investigation since accurate prediction of NO formation and extinction through the exchange process plays an important role in non-equilibrium hypersonic flow and radiation modeling \cite{Jo_2019_PRE,Jo_2020_HMT}. Figure \ref{fig:GlobalN2O_NOMole} shows the evolution of the global Zel'dovich reaction rate coefficient $\tilde{k}^{E,\text{N}_2}$ and NO mole fraction at the three different $T$. In the considered temperature range, the peak of $\chi_{\text{NO}}$ occurs before the onset of the molecular QSS periods. Moreover, at the time instants at which the NO formation starts to be relevant, the magnitude of $\tilde{k}^{E,\text{N}_2}$ is between 3 and 7 times smaller than the corresponding QSS value. Given the fact that the majority of multi-dimensional hypersonic CFD codes employ QSS rate coefficients to model the Zel'dovich reaction, the results in Figs. \ref{fig:Exch_Region_All}, \ref{fig:NoQSS_withoutDiss}, and \ref{fig:GlobalN2O_NOMole} imply that further investigation on the reliability of such an approximation is required, as it was found to be not valid for the exchange process.

\begin{figure}[h!]
    \centering
    \subfigure[]
    {
        \includegraphics[width=0.75\textwidth]{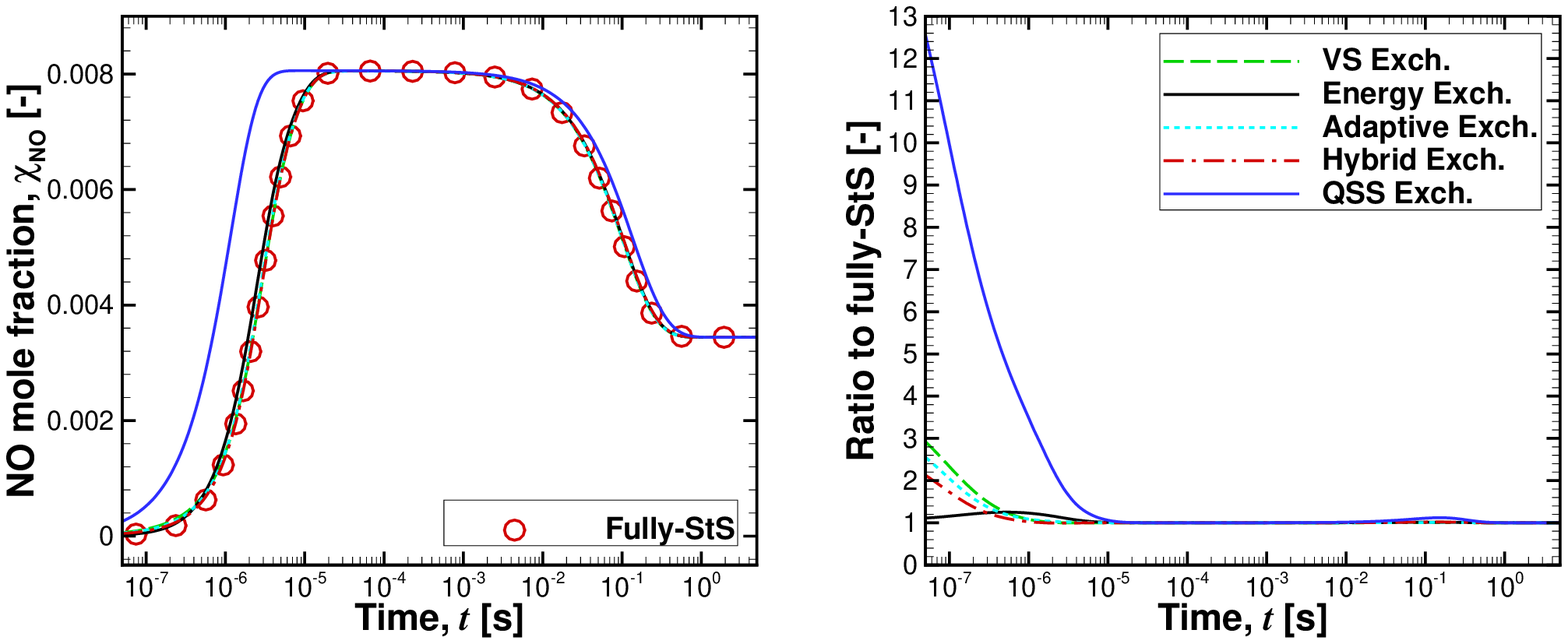}
        \label{fig:NO_T5000K}
    }
    \subfigure[]
    {
        \includegraphics[width=0.75\textwidth]{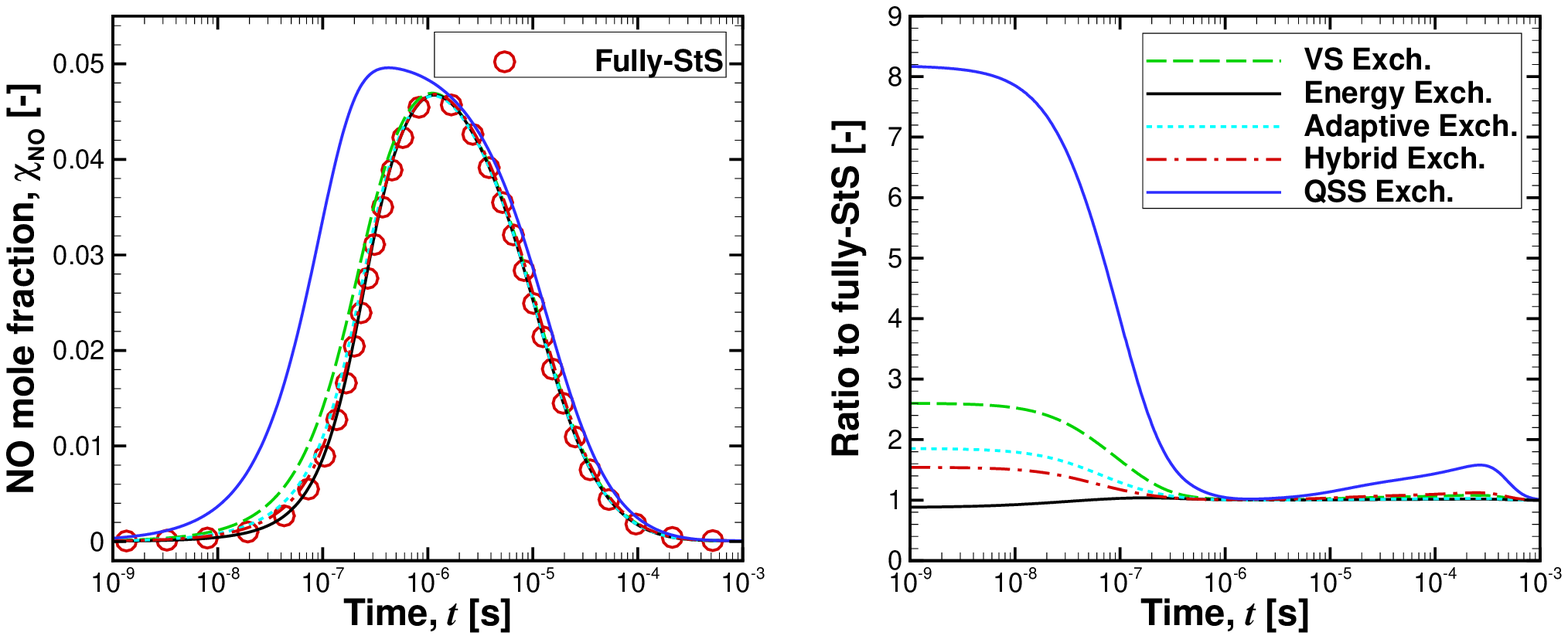}
        \label{fig:NO_T10000K}
    }
    \subfigure[]
    {
        \includegraphics[width=0.75\textwidth]{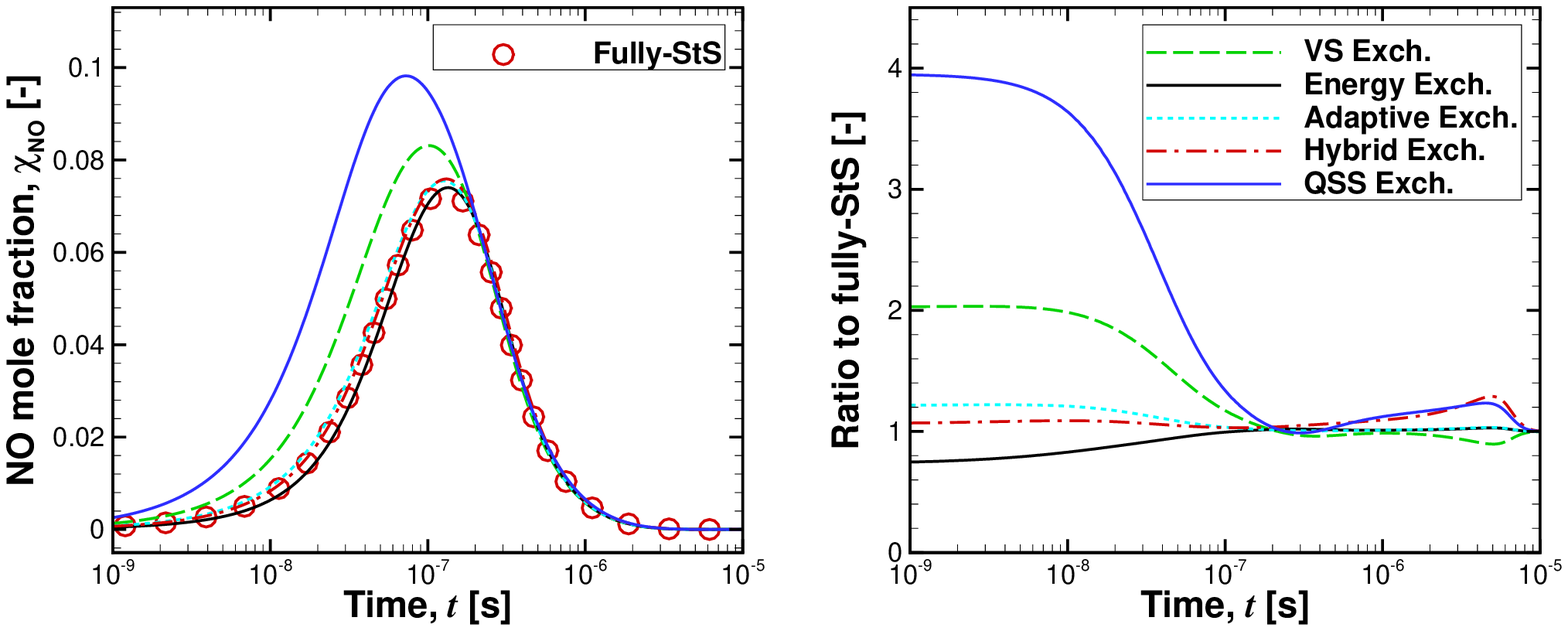}
        \label{fig:NO_T20000K}
    }
  \caption{Distributions of NO mole fraction (left). The ratio of NO mole fractions by the grouped-reconstructed methods to the fully-StS approach (right). The left and right figures share the plot legend. (a) $T=\SI{5000}{\kelvin}$, (b) $T=\SI{10000}{\kelvin}$, (c) $T=\SI{20000}{\kelvin}$. As initial mole fractions, $\chi_{\text{N}_2}$=$\chi_{\text{O}}$=0.5 is considered.}
    \label{fig:NOFormation}
\end{figure}

The accuracy of the QSS-approximated modeling for the Zel'dovich reaction can be investigated by comparing with the fully-StS results. For this purpose, the grouped-reconstructed method in Eq. (\ref{eq:step-4}) is employed to integrate the set of master equations with several reduced-order models, including the QSS approximation for the heterogeneous exchange process. Figure \ref{fig:NOFormation} shows the NO mole fraction distributions for the fully-StS (\emph{i.e.}, rovibrational state-specific) and the grouped-reconstructed method for the Zel'dovich mechanism. As the grouped-reconstructed methods, the vibration-specific (VS) \cite{Munafo_EPJD_2012}, the energy-based group (Energy) \cite{MACDONALD_MEQCT}, the adaptive coarse graining (Adaptive) \cite{SAHAI_ADAPTIVE}, and a hybrid model of the adaptive \cite{SAHAI_ADAPTIVE} and the energy deficit based \cite{Venturi2020} group methods are employed together with the QSS model. It should be mentioned that all the reduced order models employed here are based on the idea of clustering the levels in groups. Levels in the same group are assumed in local equilibrium between each other. What differentiates the reduced order models is: (1) the number of groups. The QSS model has one single group, while the remaining ones have a number of clusters equal to the amount of distinct vibrational states in the molecule. (2) The strategy adopted for clustering the levels. The macroscopic QSS rate coefficient obtained from the master equation analysis (\emph{e.g}, the empty dots in Fig. \ref{fig:GlobalN2O_NOMole}) is utilized to construct the QSS-assumption-based grouped-reconstructed rate coefficients in Eq. (\ref{eq:rate_Recon}). It should be noted that the other kinetic processes, except for the Zel'dovich reaction, remain in the resolution of fully-StS for all of the grouped-reconstructed computations. In the considered temperature range, the QSS method overpredicts the level of $\chi_{\text{NO}}$, especially in the NO formation phase and the peak value. In hypersonic non-equilibrium flow-radiation computations, the predictive accuracy on the population of electronic ground state NO affects the calculation of radiative emission profile by NO bands, which strongly contribute to the radiative heat flux. As a consequence, the overestimation of NO mole fraction by the QSS method could cause overprediction of the NO emission profile. The remaining grouped-reconstructed calculations reasonably reproduce the fully-StS profiles, except for the VS model at $T=\SI{20000}{\kelvin}$. Among the different grouping strategies, the energy-based group and the hybrid approach present better accuracy than the others. The results shown in Fig. \ref{fig:NOFormation} imply that the existing reduced order models \cite{Munafo_EPJD_2012,MACDONALD_MEQCT,SAHAI_ADAPTIVE,Venturi2020} can be employed for better description of the collisional heterogeneous exchange process of $\text{N}_2$+O $\leftrightarrow$ NO+N, instead of the QSS approximation \cite{Park_book}.

\subsection{Comparison of reaction rate coefficients}\label{sec:rate}
\begin{figure}[h]
    \centering
    \subfigure[]
    {
        \includegraphics[width=0.48\textwidth]{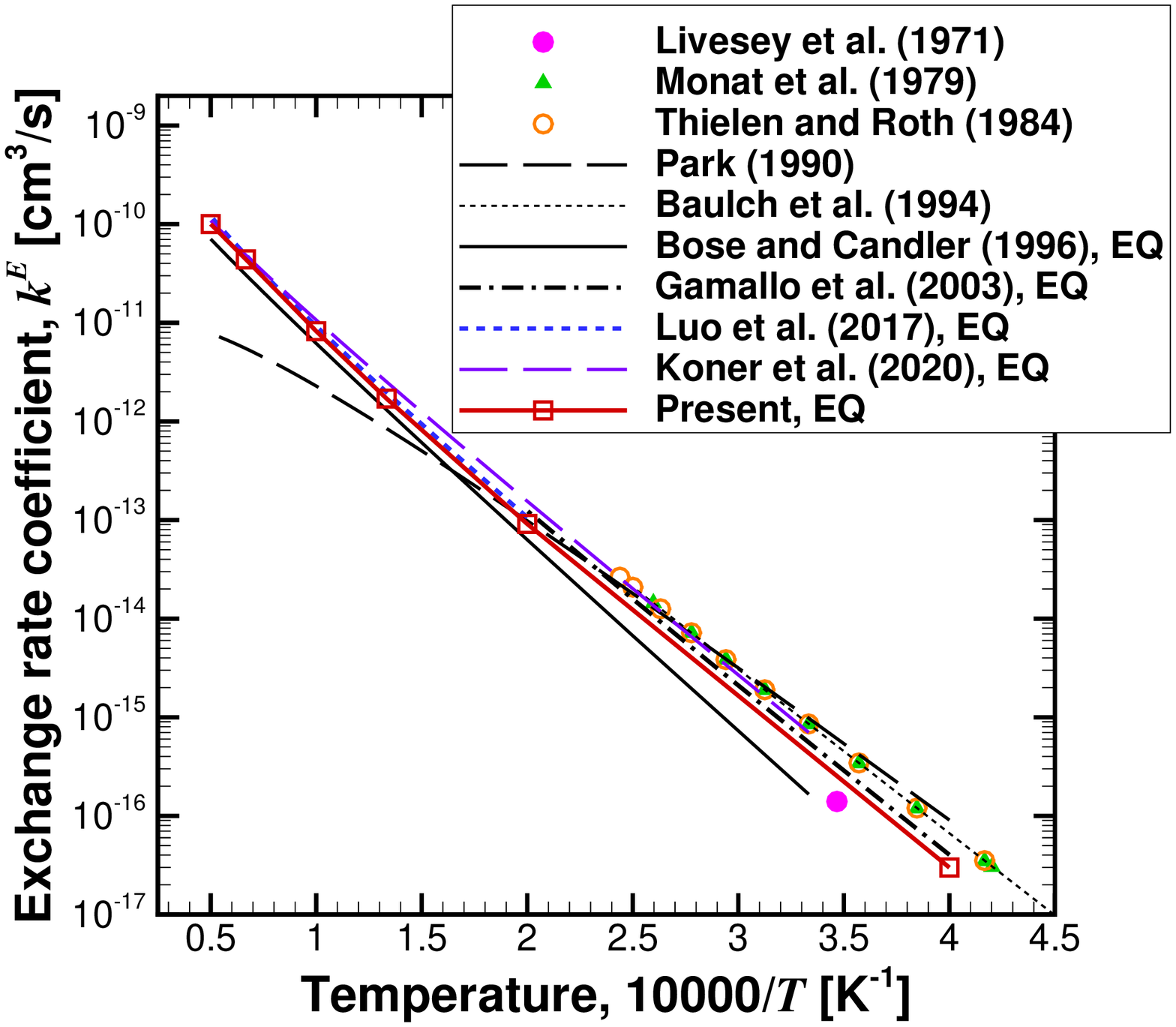}
        \label{fig:N2-O_Exch}
    }
    \subfigure[]
    {
        \includegraphics[width=0.48\textwidth]{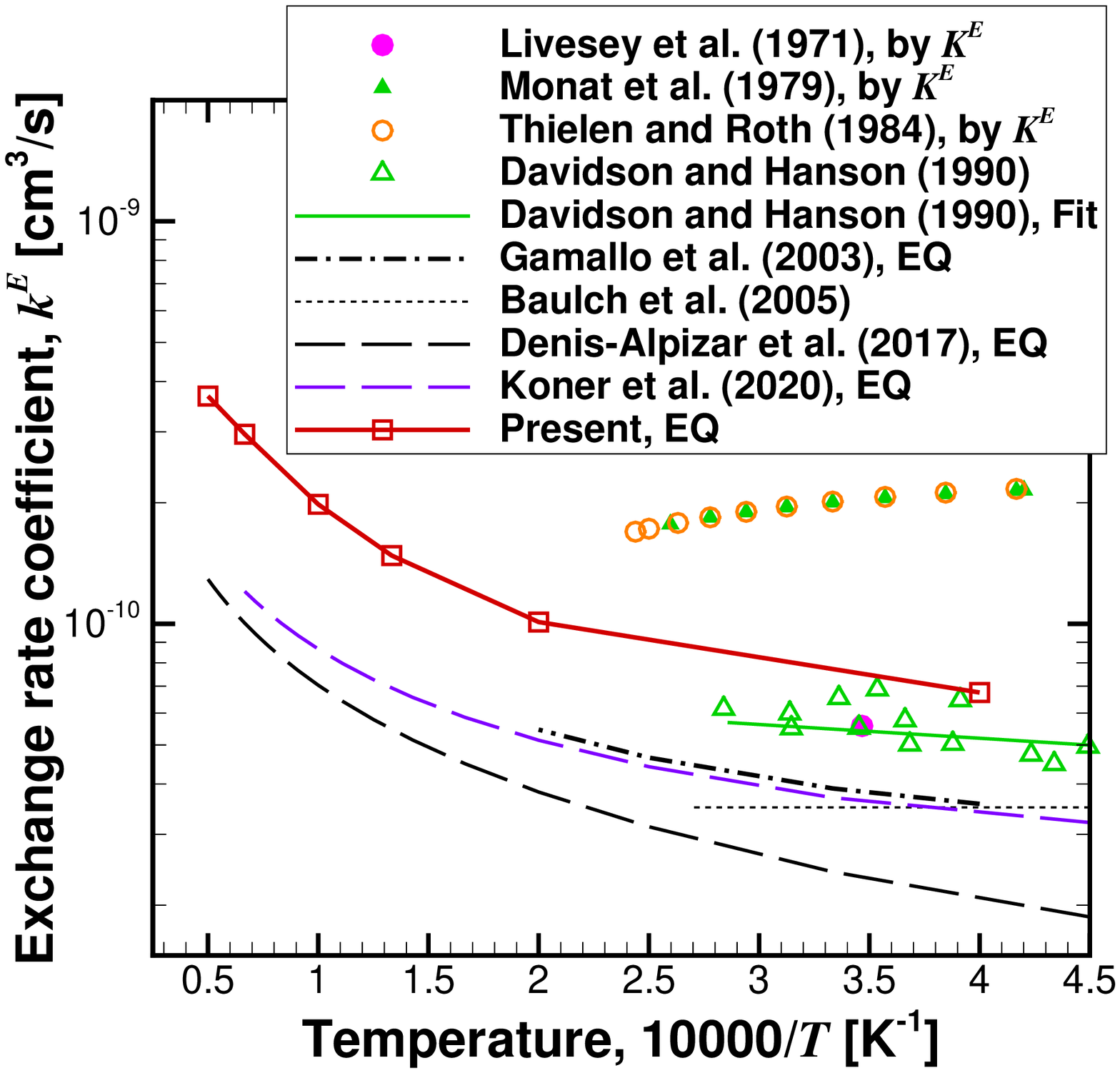}
        \label{fig:NO-N_Exch}
    }
  \caption{Comparisons of the heterogeneous exchange rate coefficients with existing experimental\cite{Livesey_1971,MONAT1979,THIELEN1985,Davidson_1990} and theoretical\cite{Park_book,Baulch_1994,Bose_JCP_1996,Gamallo_2003,Baulch_2005,Luo_2017,DenisAlpizar_NON,Koner_2020} data: (a) $\mathrm{N}_2$+O $\rightarrow$ NO+N, (b) NO+N $\rightarrow$ $\text{N}_2$+O.}
    \label{fig:ExchRate}
\end{figure}

In this section, the macroscopic reaction rate coefficients, obtained from the present QCT and master equation analyses, are compared with existing data. Figure \ref{fig:ExchRate} compares the present heterogeneous exchange rate coefficients with data in the literature \cite{Livesey_1971,MONAT1979,THIELEN1985,Davidson_1990,Park_book,Baulch_1994,Bose_JCP_1996,Gamallo_2003,Baulch_2005,Luo_2017,DenisAlpizar_NON,Koner_2020} for temperatures ranging from \SI{2500}{\kelvin} to \SI{20000}{\kelvin}. In the comparison, the QSS rate coefficient is not presented due to the following two reasons: (1) It is almost identical with the thermal rate coefficient as shown in Fig. \ref{fig:GlobalN2O_NOMole}. (2) The QSS assumption is not valid for the heterogeneous exchange, as discussed in Figs. \ref{fig:Exch_Region_All} and \ref{fig:NoQSS_withoutDiss}. 

It is worth mentioning that the present thermal rate coefficients shown in Fig. \ref{fig:ExchRate} were computed from the QCT rate coefficients corresponding to each chemical system. In detail, the QCT result for $\mathrm{N}_2$+O $\rightarrow$ NO+N was employed to calculate the thermal rate coefficient shown in Fig. \ref{fig:N2-O_Exch}, whereas the QCT results for NO+N $\rightarrow$ $\mathrm{N}_2$+O were used to obtain the results in Fig. \ref{fig:NO-N_Exch}. This enables the comparison of QCT-based rate coefficients, obtained from the different PESs \cite{Bose_JCP_1996,Gamallo_2003,Luo_2017,DenisAlpizar_NON,Koner_2020}, with the experimental data for both forward and backward directions of the Zel'dovich reaction $\mathrm{N}_2$+O $\leftrightarrow$ NO+N.

As shown in Fig. \ref{fig:N2-O_Exch}, the present thermal heterogeneous exchange rate coefficient is in reasonable agreement with the existing experimental \cite{Livesey_1971,MONAT1979,THIELEN1985} and theoretical \cite{Gamallo_2003,Luo_2017,Koner_2020} data. At temperatures over \SI{5000}{\kelvin}, the present result agrees with the thermal rate coefficients by Bose and Candler \cite{Bose_JCP_1996}, Luo \emph{et al.} \cite{Luo_2017}, and Koner \emph{et al.} \cite{Koner_2020}, whereas a departure is observed for the semi-empirically determined value by Park \cite{Park_book}. Below \SI{5000}{\kelvin}, the present result is close to the data by Gamallo \emph{et al.} \cite{Gamallo_2003}, and it is within the range of experimental error \cite{Livesey_1971,MONAT1979,THIELEN1985}.

As shown in Fig. \ref{fig:NO-N_Exch}, the heterogeneous exchange rates for NO+N $\rightarrow$ $\mathrm{N}_2$+O range over about an order of magnitude. The discrepancy between the present and the existing QCT-based results \cite{Gamallo_2003,DenisAlpizar_NON,Koner_2020} implies the uncertainty of the global triplet PESs for predicting the heterogeneous exchange kinetics through the highly exothermic reactive channel, NO+N $\rightarrow $ $\mathrm{N}_2$+O. The present result shows better agreement with the measurements by Davidson and Hanson \cite{Davidson_1990}, obtained from a resonance absorption spectrophotometry for N in a shock tube, than the QCT-based studies by Gamallo \emph{et al.} \cite{Gamallo_2003}, Denis-Alpizar \emph{et al.} \cite{DenisAlpizar_NON} and Koner \emph{et al.} \cite{Koner_2020}. Also, the current predictions are enveloped by the direct measurement \cite{Davidson_1990} and the estimated values, which are based on the experimental data \cite{MONAT1979,THIELEN1985} and the macroscopic equilibrium constant \cite{Kim_2016}, $K^E$.

\begin{figure}[h]
    \centering
    \includegraphics[width=0.55\textwidth]{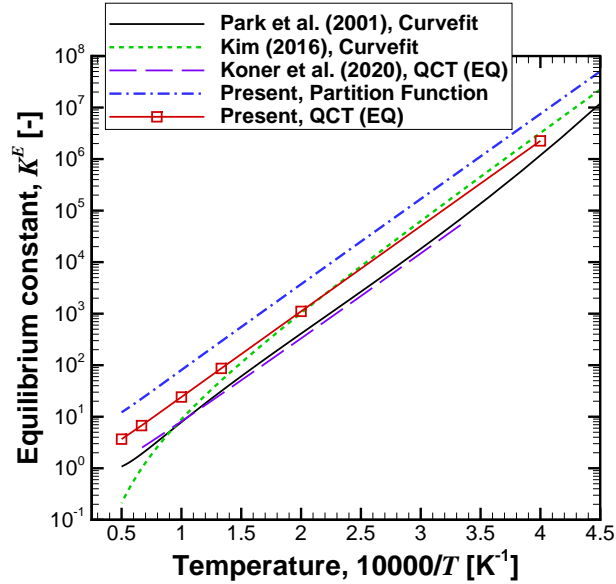}
    \caption{Comparison of the macroscopic equilibrium constant $K^E$ for $\mathrm{N}_2$+O $\leftrightarrow$ NO+N.}
    \label{fig:EQ_Const}
\end{figure}

Fidelity of the present rovibrational-specific QCT rate coefficients and the employed triplet PESs \cite{Lin_2016} are further validated by comparing the macroscopic equilibrium constant for $\mathrm{N}_2$+O $\leftrightarrow$ NO+N reaction with the existing data \cite{park2001chemical,Kim_2016,Koner_2020} and the exact values obtained based on the partition functions. Figure \ref{fig:EQ_Const} shows the equilibrium constants along the considered kinetic temperature range. For the sake of consistency with the present QCT result, the effect of excited electronic states is neglected when computing the exact equilibrium constant (\emph{i.e.}, blue dashed-dot line). On the other hand, the values from Park et al. \cite{park2001chemical} and Kim \cite{Kim_2016} contain the influence from excited electronic states. The present QCT-based equilibrium constant, obtained from the ratio of the thermal QCT rate coefficients of Fig. \ref{fig:ExchRate}, has a factor of 3 discrepancy compared to the exact value. 
This implies the leak of micro-reversibility that arises from the less accurate description of the PESs \cite{Lin_2016} for the barrierless reaction, NO+N $\rightarrow$ $\text{N}_2$+O, as addressed by Koner \emph{et al.} \cite{Koner_2020}. However, interestingly, the present QCT-based equilibrium constant shows a closer agreement with the exact value compared to the other QCT-based data by Koner et al. \cite{Koner_2020}, which has a factor of 9 discrepancy with the exact value. 

\begin{figure}[h]
    \centering
    \subfigure[]
    {
        \includegraphics[width=0.48\textwidth]{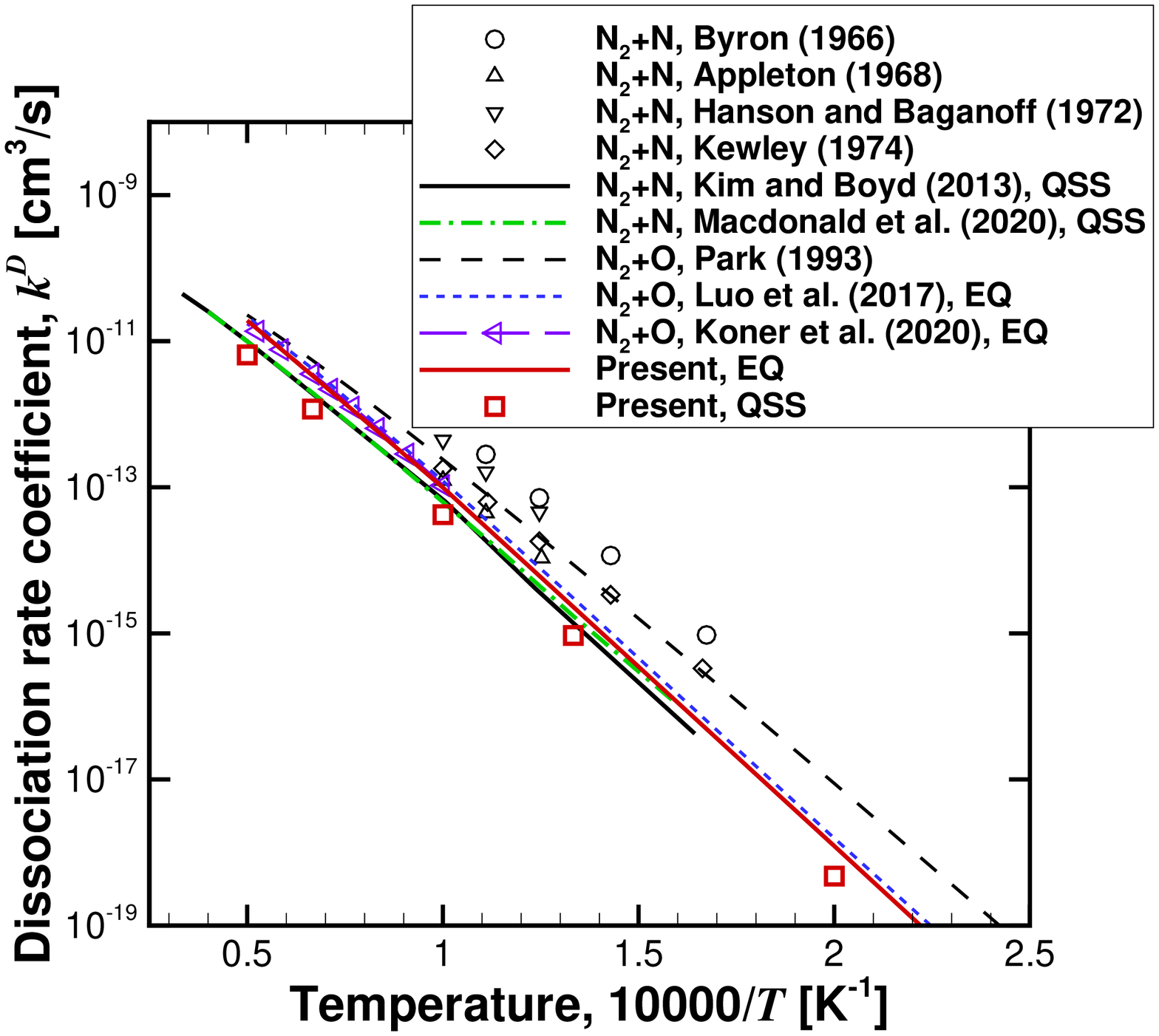}
        \label{fig:N2-O_Diss}
    }
    \subfigure[]
    {
        \includegraphics[width=0.48\textwidth]{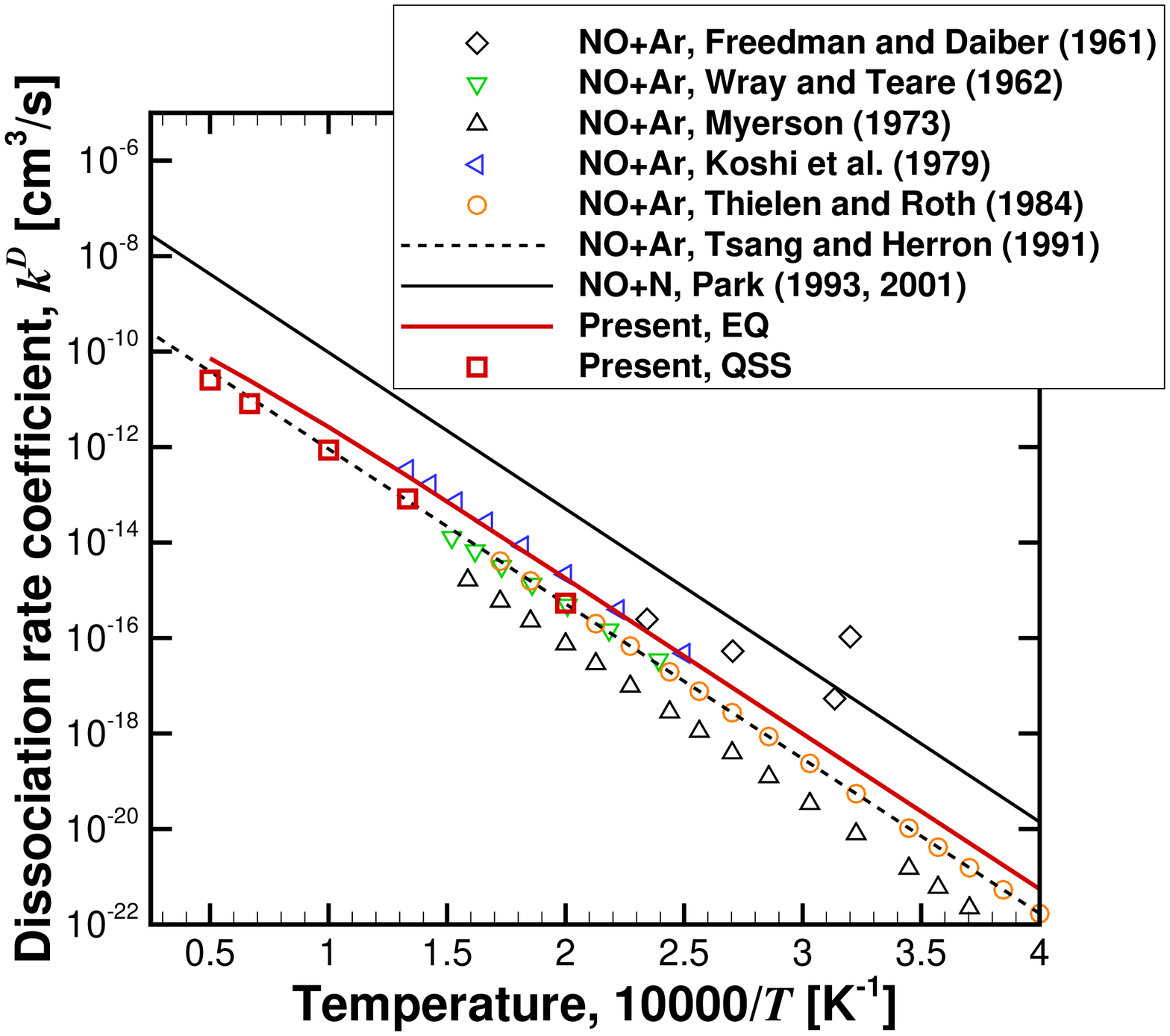}
        \label{fig:NO-N_Diss}
    }
  \caption{Comparisons of the dissociation rate coefficients with existing experimental\cite{Byron_1966,APPLETON_1968,HANSON_1972,KEWLEY1974,freedman1961,Wray1962,MYERSON1973,KOSHI1979,THIELEN1985} and theoretical\cite{Kim_Boyd_Cphys_2013,Macdonald_2020,Park_1993_Earth,Luo_2017,tsang1991chemical,park2001chemical} data: (a) $\mathrm{N}_2$+O $\rightarrow$ 2N+O, (b) NO+N $\rightarrow$ 2N+O.}
    \label{fig:DissRate}
\end{figure}

Figure \ref{fig:DissRate} compares the present dissociation rate coefficients with the existing data \cite{Byron_1966,APPLETON_1968,HANSON_1972,KEWLEY1974,freedman1961,Wray1962,MYERSON1973,KOSHI1979,THIELEN1985,Kim_Boyd_Cphys_2013,Macdonald_2020,Park_1993_Earth,Luo_2017,tsang1991chemical,park2001chemical,Koner_2020}. For both chemical systems, the QSS rate coefficients have slightly lower values than the equilibrium ones, which is consistent with the results obtained for the $\mathrm{N}_2$+N \cite{PANESI_2013_BOXRVC,Kim_Boyd_Cphys_2013} and $\mathrm{O}_2$+O \cite{DANIL_O3} systems. For the $\mathrm{N}_2$+O system, the measured $\mathrm{N}_2$+N dissociation rate coefficients \cite{Byron_1966,APPLETON_1968,HANSON_1972,KEWLEY1974} are employed in the comparison due to the lack of experimental data for the $\mathrm{N}_2$+O system. The present calculations predict lower values than the measured ones for both QSS and equilibrium. Interestingly, the present QSS dissociation rate coefficients in $\mathrm{N}_2$+O interactions are close to those for $\mathrm{N}_2$+N \cite{Macdonald_2020,Kim_Boyd_Cphys_2013}. This seems to confirm the possible similarity in physical behavior among different chemical systems that has been discussed in detail in Sec. \ref{sec:energy-iso} and Sec. \ref{sec:all-iso}. In the considered temperature range, the present equilibrium rate coefficient is in good agreement with the previous QCT results for $\mathrm{N}_2$+O by Luo \emph{et al.} \cite{Luo_2017} and Koner \emph{et al.} \cite{Koner_2020}, although the employed PESs are different. Some discrepancies appear when comparing with the data by Park \cite{Park_1993_Earth}, especially below \SI{10000}{\kelvin}. This is most probably due to the fact that the data by Park were estimated from the measurements of $\mathrm{N}_2$+N \cite{Byron_1966,APPLETON_1968,HANSON_1972}.

In order to perform comparisons for the NO+N system, measurements for NO+Ar \cite{freedman1961,Wray1962,MYERSON1973,KOSHI1979,THIELEN1985} are considered due to the lack of experimental data for dissociation in NO+N interactions. The overall spread of the measured NO+Ar dissociation rate coefficients \cite{freedman1961,Wray1962,MYERSON1973,KOSHI1979,THIELEN1985} is around four orders of magnitude. Interestingly, the QSS rate coefficient computed here are in very good agreement with the measurements by Wray \emph{et al.},\cite{Wray1962}, Thielen \emph{et al.},\cite{THIELEN1985}, and the review data by Tsang and Herron \cite{tsang1991chemical}. On the other hand, the suggested value by Park \cite{Park_1993_Earth,park2001chemical} for computing hypersonic shock layers is two orders of magnitude larger than the present calculations. In the recent study by Kim and Jo \cite{Kim_2021}, it was found that the NO dissociation rate coefficient critically affects the prediction of non-equilibrium air radiation for earth re-entry conditions. Their study employed the NO+Ar dissociation rate coefficient recommended by Tsang and Herron \cite{tsang1991chemical} as dissociation rates for NO+N and NO+O, instead of the Park value \cite{Park_1993_Earth,park2001chemical}. This fact improved the accuracy of the non-equilibrium radiation prediction, and the measured NO radiation bands \cite{Cruden2020} were better reproduced. These findings are in line with the present QCT-based dissociation rate coefficient calculations for the NO+N system shown in Fig. \ref{fig:NO-N_Diss}.

\section{Conclusions}\label{sec:conc}
This work has performed the rovibrational-specific QCT and master equation analysis of the kinetics of $\text{N}_2(\text{X}^1\Sigma_g^+)$+O$({}^3\text{P})$ and NO$(\text{X}^2\Pi)$+N$({}^4\text{S})$ systems at conditions of interest to hypersonic applications. The complete sets of the StS kinetic database on the N$_2$O system allow the investigation of detailed chemical-kinetic processes, including the Zel’dovich mechanism, at both microscopic and macroscopic scales. 
The phenomenological rate coefficients for the Zel’dovich and dissociation reactions derived in this analysis agree with available data from the literature. In contrast, marked differences are observed when comparing the results of the \emph{ab-initio} calculations with the widely adopted reaction rate coefficients (\emph{i.e.}, the Park model).
The relaxation rate parameters describing rovibrational energy transfer demonstrate the inadequacy of the correlation formulas often used in practical applications. 
Comparing the dynamics of relaxation among the N$_2$+O, NO+N, and N$_2$+N systems, investigated in detail by examining the time evolution of their rovibrational state population dynamics, reveals important similarities in the evolution of their kinetics. The three systems show a similar strand-like structure of the low-lying rovibrational states and similar evolution of the internal energy (\emph{i.e.}, energy transfer-chemistry coupling) during dissociation. This is an important finding as it facilitates the construction of reduced-order models for air chemistry. It is also found that the formation of NO and the corresponding extinction of $\text{N}_2$ through the Zel’dovich mechanism are dominated by the low-lying vibrational states. Furthermore, the distribution of the $\text{N}_2$ molecules does not reach any QSS conditions without dissociation. These findings demonstrate the invalidity of the QSS approximation widely adopted when modeling the chemical reactions in hypersonic flow simulations. 
A coarse grained description of the Zel'dovich reaction is proposed to overcome the limitation of the QSS approach. This allows to accurately describe the details of the thermochemical relaxation and the concentration profiles driven by the exchange process.

\section*{Acknowledgments} The work was supported by ONR Grant No. N00014-21-1-2475 with Dr. Eric Marineau as Program Manager. The authors would like to thank Dr. R.L. Jaffe  (NASA AMES Research Center) for the useful discussions about QCT calculations.

\appendix
\section{Mass production rates}\label{app:rates}
\begin{IEEEeqnarray}{rCl}
\dot{\omega}_i^I&=&\sum_j^{\text{NO}}k_{i{\rightarrow}j}^{\text{NO}}\left[\frac{1}{K_{i,j}^I}n_j-n_i\right]n_{\text{N}}, \label{eq:NO_Source_EnergyTransfer} \\
\dot{\omega}_m^I&=&\sum_l^{\text{N}_2}k_{m{\rightarrow}l}^{\text{N}_2}\left[\frac{1}{K_{m,l}^I}n_l-n_m\right]n_{\text{O}}, \label{eq:N2_Source_EnergyTransfer} \\
\dot{\omega}_{i,m}^{E,\text{NO}}&=&k_{i{\rightarrow}m}^{E,\text{NO}}\left[n_in_{\text{N}}-\frac{1}{K_{i,m}^E}n_mn_{\text{O}}\right], \label{eq:Source_HeteroExch} \\
\dot{\omega}_i^D&=&k_{i{\rightarrow}c}^{\text{NO}}\left[n_i-\frac{1}{K_i^D}n_{\text{N}}n_{\text{O}}\right]n_{\text{N}}, \label{eq:NO_Source_Diss} \\
\dot{\omega}_m^D&=&k_{m{\rightarrow}c}^{\text{N}_2}\left[n_m-\frac{1}{K_m^D}n_{\text{N}}^2\right]n_{\text{O}}, \label{eq:N2_Source_Diss}
\end{IEEEeqnarray}
\noindent
where $K$ stands for the equilibrium constant of a particular collisional process to evaluate reverse rate coefficients by invoking the micro-reversibility. For the collisional excitation process between $m$-th and $l$-th states of $\text{N}_2$, the equilibrium constant $K_{m,l}^I$ is defined as
\begin{equation}
K_{m,l}^I=\frac{g_l\exp\left(-\frac{e_l}{k_BT}\right)}{g_m\exp\left(-\frac{e_m}{k_BT}\right)}.
\label{eq:N2_EQConst_EnergyTransfer}
\end{equation}
\noindent
The degeneracy of the level, $g_m$, can be denoted as
\begin{equation}
g_m=g_e^{\text{N}_2}g_{nuc}^{\text{N}_2}\left(2J\left(m\right)+1\right),
\label{eq:N2_Degeneracy}
\end{equation}
\noindent
with $g_e$ and $g_{nuc}$ being the electronic ground and the nuclear spin degeneracy, respectively. $J\left(m\right)$ is the rotational quantum number of the $m$-th state. For $\text{N}_2(\text{X}^1\Sigma_g^+)$, $g_{nuc}$ is correspondingly set to 6 and 3 for the even and odd $J$ states. For the collisional heterogeneous exchange process between $i$-th and $m$-th states of NO and $\text{N}_2$, the equilibrium constant $K_{i,m}^E$ is expressed as
\begin{equation}
K_{i,m}^E=\frac{g_m\exp\left(-\frac{e_m}{k_BT}\right)Q_t^{\text{N}_2}g_e^{\text{O}}\exp\left(-\frac{e^{\text{O}}}{k_BT}\right)Q_t^{\text{O}}}{g_i\exp\left(-\frac{e_i}{k_BT}\right) Q_t^{\text{NO}}g_e^{\text{N}}Q_t^{\text{N}}}\exp\left(-\frac{E_{\text{NO}}^D - E_{\text{N}_2}^D}{k_BT}\right),
\label{eq:EQConst_HeteroExch}
\end{equation}
\noindent
where $Q_t$ is the translational partition function, and $E^D$ is the species average dissociation energy. $g_e^{\text{N}}$ and $g_e^{\text{O}}$ are the electronic degeneracy of N and O at their ground states. For the collisional dissociation process of $m$-th state of $\text{N}_2$, the equilibrium constant $K_m^D$ is defined as
\begin{equation}
K_{m}^D=\frac{\left(g_e^{\text{N}}Q_t^{\text{N}}\right)^2}{g_mQ_t^{\text{N}_2}}\exp\left(-\frac{E_{\text{N}_2}^D-e_m}{k_BT}\right).
\label{eq:N2_EQConst_Diss}
\end{equation}
\bibliography{ref}
\end{document}